\documentstyle[emulateapj]{article}

\submitted{to appear in the Astrophysical Journal, 521, No.2}

\lefthead{Hachisu, Kato, and Nomoto}
\righthead{Progenitor of Type Ia Supernovae}

\begin{document}

\title{A WIDE SYMBIOTIC CHANNEL TO TYPE Ia SUPERNOVAE}

\author{Izumi Hachisu}
\affil{Department of Earth Science and Astronomy, 
College of Arts and Sciences, University of Tokyo,
Komaba, Meguro-ku, Tokyo 153-8902, Japan \\ e-mail: 
hachisu@chianti.c.u-tokyo.ac.jp}

\author{Mariko Kato}
\affil{Department of Astronomy, Keio University, 
Hiyoshi, Kouhoku-ku, Yokohama 223-8521, Japan \\ e-mail: 
mariko@educ.cc.keio.ac.jp}

\and

\author{Ken'ichi Nomoto}
\affil{Department of Astronomy, University of Tokyo, 
Bunkyo-ku, Tokyo 113-0033, Japan \\ e-mail: 
nomoto@astron.s.u-tokyo.ac.jp\\
Research Center for the Early Universe, University of Tokyo, Bunkyo-ku,
Tokyo 113-0033, Japan}




\begin{abstract}
As a promising channel to Type Ia supernovae (SNe Ia), we have
proposed a symbiotic binary system consisting of a white dwarf (WD) and
a low mass red-giant (RG), where strong winds from the accreting WD
play a key role to increase the WD mass to the Chandrasekhar mass
limit.  However, the occurrence frequency of SNe Ia through this
channel has been still controversial.  Here we propose two new
evolutionary processes which make the symbiotic channel to SNe Ia much
wider.  (1) We first show that the WD + RG close binary can form from
a wide binary even with such a large initial separation as $a_i
\lesssim 40000 R_\odot$.  Such a binary consists of an AGB star and a
low mass main-sequence (MS) star, where the AGB star is undergoing
superwind before becoming a WD.  If the superwind at the end of AGB
evolution is as fast as or slower than the orbital velocity, the wind
outflowing from the system takes away the orbital angular momentum
effectively.  As a result the wide binary shrinks greatly to become a
close binary.  Then the AGB star undergoes a common envelope (CE)
evolution.  After the CE evolution, the binary becomes a
pair of a carbon-oxygen WD and the MS star.  When the MS star evolves
to a RG, a WD + RG system is formed.  Therefore, the WD + RG binary
can form from much wider binaries than our earlier estimate which is
constrained by $a_i \lesssim 1500 R_\odot$.  (2) When the RG fills its
inner critical Roche lobe, the WD undergoes rapid mass accretion and
blows a strong optically thick wind.  Our earlier analysis has shown
that the mass transfer is stabilized by this wind only when the mass
ratio of RG/WD is smaller than 1.15.  Our new finding is that the WD
wind can strip mass from the RG envelope, which could be efficient
enough to stabilize the mass transfer even if the RG/WD mass ratio
exceeds 1.15.  If this mass-stripping effect is strong enough, though
its efficiency is subject to uncertainties, the symbiotic channel can
produce SNe Ia for a much (ten times or more) wider range of the
binary parameters than our earlier estimation.  With the above two new
effects (1) and (2), the symbiotic channel can account for the
inferred rate of SNe Ia in our Galaxy.  The immediate progenitor
binaries in this symbiotic channel to SNe Ia may be observed as
symbiotic stars, luminous supersoft X-ray sources, or recurrent novae
like T CrB or RS Oph, depending on the wind status.

\end{abstract}




%

\section{INTRODUCTION}

      It is widely accepted that Type Ia supernovae (SNe Ia) are
thermonuclear explosions of accreting white dwarfs (e.g., Nomoto,
Iwamoto, \& Kishimoto 1997).  However, whether the explosion of the
white dwarf takes at the Chandrasekhar mass limit or at the
sub-Chandrasekhar mass has been controversial (e.g., Nomoto et
al. 1994; Arnett 1996; Branch 1998).  Also the issue of double
degenerate (DD) vs. single degenerate (SD) for the progenitor scenario
has been still debated (e.g., \cite{bra95} for a review).

      The DD scenario assumes that merging of double C+O white dwarfs
with a combined mass surpassing the Chandrasekhar mass limit induces
SN Ia (e.g., \cite{ibe84}; \cite{web84}).  However, this scenario has
not been well supported.  Observationally, the search for DDs has
discovered only several systems whose combined mass is less than the
Chandrasekhar mass or whose separation is too wide to merge in a
Hubble time (\cite{bra95}; \cite{ren96}; \cite{liv96} for reviews).
Theoretically, the DD has been suggested to lead to
accretion-induced-collapse rather than SN Ia (\cite{nom85};
\cite{sai85}, 1998; \cite{seg97}).

     For the SD scenario, observational counterparts may be symbiotic
stars (e.g., \cite{mun92}).  Kenyon et al. (1993) and Renzini (1996)
have suggested that symbiotics are more likely to lead to the
sub-Chandrasekhar mass explosion because the available mass in
transfer may not be enough for white dwarfs to reach the Chandrasekhar
mass.  However, photometric and spectroscopic features of majority of
SNe Ia are in much better agreement with the Chandrasekhar mass model
than the sub-Chandrasekhar mass model (\cite{hof96}; \cite{nug97}).

     Hachisu, Kato, \& Nomoto (1996; hereafter HKN96) have shed new
light on the above SD/sub-Chandrasekhar mass scenario and proposed a new
progenitor system based on the optically thick wind theory of
mass-accreting white dwarfs.  In the scenario of Iben \& Tutukov
(1984) and Webbink (1984), they excluded close binary systems
consisting of a mass-accreting white dwarf (WD) and a lobe-filling
red(sub)-giant (RG) (first discussed by \cite{whe73}), mainly because
such a system suffers from unstable mass transfer when the mass ratio
of the RG/WD exceeds 0.79, i.e., $q=M_{\rm RG}/M_{\rm WD} > 0.79$.
However, HKN96 have shown that optically thick winds from the
mass-accreting white dwarf stabilize the mass transfer up to $q \le
1.15$ even if the RG has a deep convective envelope.  Such an object
may be observed as a symbiotic star.  Thus they proposed a new channel
to SNe Ia, through which a white dwarf accreting mass from
a lobe-filling red-giant can grow up to the Chandrasekhar mass limit
and explodes as an SN Ia (SD/Chandrasekhar mass scenario of symbiotics). 
\par
     Li \& van den Heuvel (1997) reanalyzed the HKN96 model 
and identified two isolated regions for SN Ia progenitors 
in the initial orbital period vs. the initial donor 
mass plane, i.e., in the $\log P_0 - M_{\rm d,0}$ plane:  
1) One is a relatively compact close binary consisting
of a $M_{\rm d,0} \sim 2-3 M_\odot$ slightly evolved main-sequence 
companion and a $M_{\rm WD,0} \sim 1.0-1.2 M_\odot$ 
white dwarf with the initial orbital periods 
of $P_0 \sim 0.5-5$ d (hereafter, WD+MS systems), 
and 2) the other is a relatively wide binary consisting of a low mass 
($M_{\rm d,0} \sim 1 M_\odot$) red-giant companion
and a $M_{\rm WD,0} \sim 1.2 M_\odot$ white dwarf 
with the initial orbital periods of $P_0 \sim 100-800$ d 
(hereafter, WD+RG systems).  
They also concluded that the new model 
accounts for the inferred rate of SNe Ia in our Galaxy.
\par
     In Li \& van den Heuvel's (1997) analysis,
their SN Ia progenitor region for the WD+RG system is very small 
compared with the region for the WD+MS system.  
The contribution of the WD+RG systems to the total rate 
of SN Ia explosions was expected to be very small 
or negligible.  It is because the WD+RG channel 
is restricted by the condition for the stable mass transfer, 
i.e., $q < 1.15$.  In the present paper, 
we propose a new evolutionary process which makes 
the WD+RG channel 
to SN Ia much wider than that of HKN96's original modeling.   
We include a mass-stripping effect 
of the red-giant envelope by the wind.
The effect removes the limitation of $q < 1.15$ and, as a result, 
the new parameter region producing an SN Ia becomes ten times or more 
wider than the previous region calculated 
by HKN96 and Li \& van den Heuvel (1997).  
\par
     Recently, Yungelson \& Livio (1998) claimed, 
based on their population synthesis results, 
that HKN96's  and Li \& van den Heuvel's 
(1997) model can account for only, at most, 
10\% of the inferred rate of SNe Ia. 
Introducing a new evolutionary process into HKN96's modeling,  
we have reanalyzed the SN Ia rate for our extended HKN96's model 
and also for Li \& van den Heuvel's model.
Our present analysis reveals that the realization frequencies of 
SNe Ia coming from our WD+RG/WD+MS models are 
$\sim 0.002$ yr$^{-1}$/$\sim 0.001$ yr$^{-1}$, 
respectively, and the total SN Ia rate 
becomes $\sim 0.003$ yr$^{-1}$, which is large enough to account 
for the inferred rate of SN Ia rate in our Galaxy.
There are three reasons why our estimates are much larger 
than Yungelson \& Livio's (1998) estimates: 
1) We introduce a mass-stripping process into the WD+RG systems
as an extension of HKN96's model.  As a result, 
the new parameter region producing an SN Ia becomes ten times or more 
wider than the previous region calculated 
by HKN96 and Li \& van den Heuvel (1997) when the efficiency 
of the mass-stripping effect is strong enough.  
2) Yungelson \& Livio (1998) assumed that the initial separation is 
{\it smaller than} $a_i \lesssim 1500 ~R_\odot$ 
in their estimations of the WD+RG model.  
If one includes the effect of angular momentum loss by slow winds 
at the end of stellar evolution, however, very wide binaries with 
the separation of $1500 ~R_\odot \lesssim a_i \lesssim 40000 ~R_\odot$ 
shrink into $30 ~R_\odot \lesssim a_f \lesssim 800 ~R_\odot$, 
which provide appropriate initial conditions for our WD+RG models.
3) We believe that Yungelson \& Livio (1998) did not include 
an important evolutionary path in the rate estimation for 
Li \& van den Heuvel's (1997) WD+MS model.  
They assumed that relatively massive white dwarfs 
($\sim 1 M_\odot$) are born from an AGB star and neglected 
the possibility that it comes from a helium star whose 
hydrogen-rich envelope is stripped away in a common envelope 
evolution at the red-giant phase with a helium core.  
Very recently, including the evolutionary path mentioned above,
Hachisu, Kato, Nomoto, \& Umeda (1999, hereafter HKNU99) 
have shown that the realization frequency for the WD+MS systems 
is as large as $\nu_{\rm MS} \sim 0.001$ yr$^{-1}$, which accounts for 
one third of the inferred rate of SNe Ia in our Galaxy.
\par
     It has been argued that some of the recurrent novae are 
progenitors of SNe Ia (e.g., \cite{sta88}) 
because these white dwarfs are suggested 
to be very massive and close to the Chandrasekhar mass limit. 
Morphologically, recurrent novae are divided into three groups
according to the companion star; 
dwarf companions, slightly evolved main-sequence 
(or sub-giant) companions, and red-giant companions (\cite{sch95}).
The latter two groups are relevant to SNe Ia progenitors and
the good examples are as follows.
\par
     1) T CrB ($P_{\rm orb}= 227.67$ d; e.g., \cite{lin88}) 
and RS Oph ($P_{\rm orb}= 460$ d; \cite{dob94})
belong to the last group of red-giant companions and 
their white dwarf masses are very close to 
the Chandrasekhar mass limit (e.g., \cite{kat90}, 1995, 1999;
\cite{sha97}; \cite{bel98}; \cite{hac99ka}). 
These two systems correspond to the WD+RG systems. 
These extremely massive white dwarfs are naturally explained 
in our SN Ia progenitor scenario.
\par
     2) On the other hand, U Sco 
($P_{\rm orb}= 1.23$ d; \cite{sch95}) 
and V394 CrA ($P_{\rm orb}= 0.758$ d; \cite{sch90}) 
belong to the middle group of the slightly evolved main-sequence 
companions.  For this group, it has been suggested that the companion
has an extremely helium-rich envelopes and the primary is a 
very massive white dwarfs close to the Chandrasekhar mass limit. 
Our WD+MS model yields the secondary star having 
a helium-rich envelope 
as well as the primary of very massive white dwarfs 
as suggested first by Hachisu \& Kato (1999b).
\par 
     In \S 2, we describe a new idea of mass-stripping effect 
by strong winds and the mass accumulation efficiency.  We then  
search for the initial parameter regions that can produce 
SNe Ia in \S 3.  
In \S 4, we discuss relevance to recurrent novae, our criticism to 
Yungelson \& Livio's claims, estimation of SN Ia rates 
of the WD+RG/WD+MS systems, 
and the possibility of detecting hydrogen-lines at SN Ia 
explosions during the strong wind phase.  Conclusions follow in \S 5.

\section{PROGENITOR SYSTEMS}
     First of all, we illustrate a full evolutionary path of 
our WD+RG system
from the zero age main-sequence stage ({\it stage A}) 
to the SN Ia explosion ({\it stage F}) in Figure \ref{channel}.
\begin{enumerate}
     \item[{\it A})] Zero age main-sequence.
\item[{\it B})] The primary has evolved first to become an AGB star
and blows
a slow wind (or a super wind) at the end of stellar evolution.
\item[{\it C})] The slow wind carries the orbital angular momentum
and, as a reaction, the separation shrinks considerably 
(by about a factor of ten or more), which is a similar process 
to the common envelope evolution.
\item[{\it D})] A carbon-oxygen white dwarf (the initial primary) 
and a zero age main-sequence star (the initial secondary) remain.
\item[{\it E})] The initial secondary has evolved 
to a red-giant forming
a helium core and fills up its inner critical Roche lobe.
Mass transfer begins.  The WD component blows a strong wind 
and the winds can stabilize the mass transfer even if the RG 
component has a deep convective envelope.
\item[{\it F})] The WD component has grown in mass 
to the Chandrasekhar mass
limit and explodes as a Type Ia supernova.
\end{enumerate}
\par
     For an immediate progenitor system of Type Ia supernovae (SNe Ia), 
we consider a close binary initially consisting of a
carbon-oxygen white dwarf (C+O WD) with
$M_{\rm WD,0}=  0.6-1.2 M_{\odot}$ and a low-mass
red-giant star with 
$M_{\rm RG,0}= 0.7-3.0 M_{\odot}$ having a helium core of 
$M_{\rm He,0}= 0.2-0.46  M_{\odot}$ ({\it stage E}). 
The initial state of these immediate progenitors is specified 
by three parameters, i.e., 
$M_{\rm WD,0}$, $M_{\rm RG,0}$, and the initial orbital period $P_0$
($M_{\rm He,0}$ is determined if $P_0$ is given).
We follow binary evolutions of these systems by
using empirical formulae (\cite{web83}) 
and obtain the parameter range(s) which can produce an SN Ia.  
\placefigure{channel}

\subsection{Conventional evolution scheme}
     When the companion evolves to a red-giant (RG) and
fills  its inner critical Roche lobe, mass transfer begins from
the RG to the WD.  If both the total mass and the total 
angular momentum are conserved and the mass transfer is steady, 
its rate is given by
\begin{equation}
{{\dot M_2}\over{M_2}}  =  \left({{\dot R_2}\over
{R_2}}\right)_{\rm EV} /{H(q)}, 
\end{equation}
where ${(\dot R_2/R_2)_{\rm EV}}$ represents specifically the
evolutionary change in the secondary radius and
\begin{equation}
H(q)= { {d \ln f(q)} \over {d \ln q}}(1+q) - 2(1-q),
\end{equation}
where $q$ is the mass ratio defined by
\begin{equation}
q \equiv M_2/M_1,
\end{equation}
(${M_1}$ is the mass of the primary, i.e., the WD component, 
and ${M_2}$ the mass of the secondary, i.e., the RG component).   
Here we use the empirical formula proposed by Eggleton (1983),
\begin{equation}
{R^*_2 \over a} = f(q) = { {0.49 q^{2/3} } 
\over {0.6q^{2/3}   +  \ln  (1+q^{1/3})}},
\label{empirical_roche}
\end{equation}
for an effective radius $(R^*_2)$ of the secondary's inner critical 
Roche lobe.  For the separation
$a$, we simply assume a circular orbit.   
To  estimate
${(\dot R_2/R_2)_{\rm EV}}$ we use the empirical formulae proposed by
Webbink et al. (1983).  
\par
     For a sufficiently large mass of the secondary $M_2$  (i.e.,
$q  >  0.79$), however, equation (1) gives a  positive  value  of
$\dot M_2$. This means that the mass transfer proceeds not on  an
evolutionary time scale but rather on a thermal or dynamical time
scale.  The gas falls very rapidly onto the WD and
forms an extended envelope around the WD (e.g.,  
\cite{nom79}; \cite{ibe88}).  This envelope expands  to
fill  the inner and then outer critical Roche lobe.  
It eventually results in the formation of a common envelope, 
in which the two cores are 
spiraling in each other.  It forms a very compact binary system 
consisting of a C+O WD and a helium WD, or a merger of C+O and He 
cores.  These systems have been extensively examined 
by many authors (e.g., \cite{ibe93} for a review and 
references therein). 

\subsection{White Dwarf Winds}
     However, the recent version of opacity (\cite{igl96}) 
has changed the story.  Optically thick winds are driven when
the WD envelope expands and the photospheric temperature 
decreases below $\log T_{\rm ph} \sim 5.5$ (\cite{kat94}).  
We have calculated such wind solutions for various 
white dwarf masses of $M_{\rm WD}= 0.6$, 0.7, 
0.8, 0.9, 1.0, 1.1, 1.2, 1.3, 1.35 and $1.377 M_\odot$ and
show six of ten cases, i.e, 
$M_{\rm WD}= 0.6$, 0.8, 1.0, 1.2, 1.3 and $1.377 M_\odot$
in Figures \ref{dmdtenvx70z02}$-$\ref{dmdtescx70z02}.  
Here, we choose $1.377 M_\odot$ as a limiting mass just below
the mass at the SN Ia explosion in W7 ($1.378 M_\odot$, 
\cite{nom84}) as was done in Kato (1995, 1999).
We have used the updated OPAL opacity (\cite{igl96}) 
because its strong peak near $\log T \sim 5.2$ is about $20-30$\% 
larger than that of the original OPAL opacity (\cite{rog92}) 
which was used in HKN96.  
The numerical method and various assumptions are 
the same as in Kato \& Hachisu (1994) so that we omit the details
of the numerical calculations to avoid the duplication.
The only difference between the results in HKN96 (or \cite{kat94})
and the present ones is the opacity as mentioned above.
\par
     Each wind solution is a unique function of the envelope mass
$\Delta M$ if the white dwarf mass is given.  The envelope mass 
is decreasing due to wind mass loss $\dot M_{\rm wind} ~(<0)$ 
and hydrogen shell burning $\dot M_{\rm nuc} ~(<0)$, i.e.,
\begin{equation}
{{d} \over {d t}} \Delta M = - \dot M_2 + \dot M_{\rm wind}
+ \dot M_{\rm nuc}.
\label{time_depend_envelope_mass}
\end{equation} 
When the mass transfer rate from the companion $\dot M_2 ~(<0)$ does
not vary so much in a thermal time scale of the WD envelope,
the WD envelope reaches an equilibrium of
\begin{equation}
\dot M_2 = \dot M_{\rm wind}+ \dot M_{\rm nuc}. 
\end{equation}
Thus we regard the ordinates 
in Figures \ref{dmdtenvx70z02}$-$\ref{dmdtescx70z02}
as the mass transfer rate from the companion.   
\par
     Figure \ref{dmdtenvx70z02} shows the envelope mass of 
the wind and static solutions against the total mass decreasing rate 
of the envelope $|d M/d t|$, i.e., 
the mass transfer rate from the companion $|\dot M_2|$.
There exists only a static (no wind) solution below the break 
of each solid line while there exists only a wind solution
above the break for a given envelope mass of $\Delta M$.
The optically thick winds blow when the mass transfer rate
from the companion star 
$|\dot M_2| = |\dot M_{\rm nuc} + \dot M_{\rm wind}|$ 
exceeds 
\begin{equation}
\dot M_{\rm cr} \approx 0.75 \times10^{-6} 
\left({M_{WD} \over {M_\odot}}  -
0.40\right) M_\odot {\rm ~yr}^{-1}, 
\label{critical_rate}
\end{equation}
which is reduced from our wind/static solutions as shown in Figure 
\ref{dmdtenvx70z02}.
Figure \ref{dmdttphx70z02} shows the photospheric temperature 
against the mass transfer rate.  It should be noted that
the optically thick winds begin near $\log T \sim 5.5$, which 
corresponds to the shoulder of the strong peak of 
OPAL opacity in the high temperature side.
Figure \ref{dmdtrphx70z02} depicts the photospheric radius against
the mass transfer rate.  The optically thick winds occur when 
the photosphere expands to $R_{\rm ph} \sim 0.1 R_\odot$, 
which is much smaller than the inner critical Roche lobe.
We plot the photospheric velocity $v_{\rm ph}$ 
in Figure \ref{dmdtvelx70z02} and the ratio $v_{\rm ph}/ v_{\rm esc}$ 
between the photospheric velocity and the escape velocity 
at the photosphere in Figure \ref{dmdtescx70z02} 
against the mass transfer rate.
Here, we call the wind as ``strong'' when the photospheric 
velocity exceeds the escape velocity there.  When the wind is 
strong enough, the photospheric velocity is as high as 
$\sim 1000$ km s$^{-1}$ being much faster 
than the orbital velocity, i.e., $v_{\rm ph} \gg a\Omega_{\rm orb}$.   
\placefigure{dmdtenvx70z02}
\placefigure{dmdttphx70z02}
\placefigure{dmdtrphx70z02}
\placefigure{dmdtvelx70z02}
\placefigure{dmdtescx70z02}

\par
     The optically thick wind is a continuum-radiation
driven wind in which the acceleration occurs deep inside
the photosphere (e.g., \cite{kat94}).  
Further acceleration of the wind near the photosphere is negligibly
small because the photon momentum near the photosphere is much smaller
than the momentum of the wind.  Therefore, our results are almost
independent of the treatment of the radiative transfer near the
photosphere.   In other words, 
the critical points of the wind solutions, 
where the wind is accelerated,
exist deep inside the photosphere and also deep 
inside the inner critical Roche lobe for our WD+RG systems.
This means that the binary motion does not affect 
the acceleration of the optically thick winds mainly 
because the strong winds are already accelerated deep inside
the inner critical Roche lobe up to the velocity much faster than 
the orbital motion.  
\par
     Once the strong wind 
occurs, the mass transfer rate is modified as 
\begin{equation}
{{\dot M_2}\over{M_2}}  =  \left( \left({{\dot R_2}\over
{R_2}}\right)_{\rm EV} - H_1(q) \left({{\dot M_1} \over {M_1}} 
\right) \right)/{H_2(q)}, 
\end{equation}
\begin{equation}
H_1(q)= - { {d \ln f(q)} \over {d \ln q}}   
+ {1 \over {1+q}} - 2 + 2 \ell_{\rm w} {{1+q} \over q},
\end{equation}
\begin{equation}
H_2(q)= { {d \ln f(q)} \over {d \ln q}}  
+ {q \over {1+q}} - 2 + 2 \ell_{\rm w} (1+q),
\end{equation}
where $\ell_{\rm w}$ is the specific angular momentum of the wind 
in units of $a^2 \Omega_{\rm orb}$, 
\begin{equation}
\left({{\dot J} \over {\dot M}}\right)_{\rm wind} = 
\ell_{\rm w} a^2 \Omega_{\rm orb},
\label{specific_angular_momentum_wind}
\end{equation}
$J$ the total angular momentum, $M$ the total mass of 
the system, and $\Omega_{\rm orb}$ the orbital angular velocity.  
The wind velocity is about several hundred to one thousand 
km s$^{-1}$ for relatively massive WDs when the mass transfer rate is  
$\lesssim 1 \times 10^{-5} M_\odot {\rm ~yr}^{-1}$
(Fig. \ref{dmdtvelx70z02}).  The wind 
velocity is about ten times faster than the orbital velocity 
because of $a \Omega_{\rm orb} \sim 30-100$ km s$^{-1}$ for 
$M_{\rm WD} \sim 1 M_\odot$, $M_2 \sim 1 M_\odot$, 
and $a \sim 30-400 ~R_\odot$. 
In such cases, winds cannot get angular momentum
from the orbital motion by torque during its journey, so that wind
has the same specific angular momentum 
as the WD, which is estimated as
\begin{equation}
\ell_{\rm w} = \left( {{q} \over {1+q}} \right)^2.
\label{fast_wind_angular_momentum}
\end{equation} 
In this case,
function $H_2(q)$ changes its sign at $q= 1.15$.  Wind mass loss 
stabilizes the mass transfer in the region from $q = 0.79$ to $q = 1.15$ 
(HKN96).

\subsection{Growth of white dwarfs}
      During the strong wind phase, 
as hydrogen steadily burns on the surface of the WD, 
the WD accretes the processed matter approximately at a rate of
$\dot M_{\rm cr}$, which is given in equation 
(\ref{critical_rate}).
When the mass transfer rate decreases 
below this critical value, optically thick winds stop.  If the mass 
transfer rate further decreases below 
\begin{equation}
\dot M_{\rm st} \approx {1 \over 2} \dot M_{\rm cr},
\label{steady}
\end{equation}
which is also reduced from our solutions for the minimum envelope mass
corresponding to the lower end of each solid lines in Figure
\ref{dmdtenvx70z02}, hydrogen shell burning becomes unstable to 
trigger very weak shell flashes.  Once a shell flash occurs, 
a part of the transferred hydrogen-rich matter may be blown off 
and we need to estimate the net mass accumulation in order 
to examine whether or not the WD will grow up to the Chandrasekhar 
mass limit (e.g., \cite{kov94}).  Here, we roughly assume that 
all the processed matter is retained until 
the accretion rate becomes below
\begin{equation}
\dot M_{\rm low} = {1 \over 8} \dot M_{\rm cr}
\sim 1 \times 10^{-7} M_\odot \mbox{~yr}^{-1}.
\label{accumulation}
\end{equation}
This assumption may slightly underestimate the mass loss from 
the WD, but there still exists 
a large uncertainty in relation to 
the mass accumulation ratio, $\eta_{\rm H}$, i.e., the ratio of 
the retained mass to the transferred mass after 
many cycles of hydrogen shell flashes.  To summarize, we assume
the mass accumulation ratio of hydrogen shell burning as
\begin{equation}
\eta_{\rm H} = \left\{
        \begin{array}{@{\,}l}
          0, 
\quad \mbox{~for~} |\dot M_2| \le \dot M_{\rm low} \cr
          1, 
\quad \mbox{~for~} \dot M_{\rm low} < |\dot M_2| < \dot M_{\rm cr} \cr
          1 - \dot M_{\rm w} / \dot M_2,
\mbox{~~~for~} \dot M_{\rm cr} \le |\dot M_2| \lesssim \dot M_{\rm high}
        \end{array}
      \right.
\label{hydrogen_accumulation}
\end{equation}
where $\dot M_{\rm w}$ is the wind mass loss rate calculated by
the optically thick wind theory and
$\dot M_{\rm high} \sim 1 \times 10^{-4} - 1 \times 10^{-3} 
M_\odot$ yr$^{-1}$ is the upper limit of the mass transfer rate 
for our wind solutions as shown in Figure \ref{dmdtenvx70z02}.  
\par
     The steady hydrogen shell burning converts hydrogen into
helium atop the C+O core and increases the mass of the helium layer
gradually.   When its mass reaches a certain value,
helium ignites.   For the accretion rate given by equation
(\ref{critical_rate}), 
helium  shell  burning is unstable to grow to a weak flash.   Once  a
helium  shell  flash occurs on relatively  massive  white  dwarfs
($M_{\rm WD}  \gtrsim 1.2 M_{\odot}$), a part of the envelope mass  is
blown  off in the wind (\cite{kat89}).  
Recently, Kato \& Hachisu (1999) have recalculated the mass 
accumulation efficiency for helium shell flashes with the new opacity
(\cite{igl96}).  
Here, we adopt their new results in a simple analytic form,
\begin{equation}
\eta_{\rm He} = \left\{
        \begin{array}{@{\,}l}
          -0.175 \left( \log \dot M_{\rm He} + 5.35 \right)^2 
+ 1.05, \cr 
\qquad \mbox{~for~} -7.3 < \log \dot M_{\rm He} < -5.9 \cr
          1,   \quad \mbox{for~} -5.9 \le \log \dot M_{\rm He} 
\lesssim -5
        \end{array}
      \right.
\label{helium_accumulation}
\end{equation}  
where the mass accretion rate,
$\dot M_{\rm He}$, is in units of $M_\odot$ yr$^{-1}$ and given by
\begin{equation}
\dot M_{\rm He} = \eta_{\rm H} \cdot |\dot M_2|,
\end{equation}
when hydrogen-rich matter is transferred from the companion star.
Eventually, we have the growth rate of C+O WDs given by
\begin{equation}
\dot M_{\rm C+O} = \eta_{\rm He} \cdot \dot M_{\rm He}
= \eta_{\rm He} \cdot \eta_{\rm H} \cdot |\dot M_2|,
\end{equation}
in our WD+RG systems.
We use formula (\ref{helium_accumulation})
for various white dwarf masses and accretion rates,
although their results are given only for 
a $1.3 M_\odot$ white dwarf (\cite{kat99h}). 
\par
     The wind velocity in helium shell flashes reaches 
as high as $\sim 1000$ km s$^{-1}$,
which is much faster than the orbital velocities of our
WD+MS binary systems, i.e., 
$a\Omega_{\rm orb} \sim 300$ km s$^{-1}$
or of our WD+RG binary systems, i.e., 
$a\Omega_{\rm orb} \sim 30$ km s$^{-1}$.
It should be noted here that either a Roche lobe overflow or a  
common envelope does not play a role as a mass ejection mechanism
because the envelope matter goes away quickly from the system 
without interacting with the orbital motion 
(see \cite{kat99h} for more details).

\placefigure{collision}

\subsection{Mass-stripping effect}
     We propose here a new effect, namely,
stripping of a red-giant envelope by the wind, 
which is not included in HKN96's modeling. 
Very fast strong winds collide with the surface of 
the companion as illustrated in Figure \ref{channel} 
({\it stage E}) and, for more detail, in Figure \ref{collision}. 
The red-giant surface is shock-heated and ablated in the wind. 
We estimate the shock-heating by 
assuming that the velocity component normal to the red-giant
surface is dissipated by shock  
and its kinetic energy is converted into the thermal energy 
of the surface layer of the red-giant envelope 
(see Fig. \ref{collision}).  
The very surface layer of the envelope expands to be easily ablated 
in the wind.  To obtain the mass stripping rate, we equate 
the rate times 
the gravitational potential at the red-giant surface to the net 
dissipation energy per unit time by the shock, i.e.,
\begin{eqnarray}
- {{G M_2} \over {R_2}} \left(d \dot M_2 \right)_{\rm strip}  
&=&  \eta_{\rm eff} \cdot 
{1 \over 2} v^2 \sin^2\beta \cdot \rho v \sin \beta 
\cr &\times&
2 \pi R_2^2 \sin\theta d \theta, 
\end{eqnarray}
between $\theta$ and $\theta+d \theta$, where 
\begin{equation}
\alpha + \beta + \theta = {\pi \over 2},
\end{equation}
\begin{equation}
{{R_2} \over {\sin \alpha}} = {{a} \over \sin ({\pi/2+\beta)}},
\end{equation}
\begin{eqnarray}
\theta &=& \cos^{-1}\left({{R_2} \over {a}} \cos \beta \right) 
- \beta \cr
&=& \cos^{-1}\left( f(q) \cdot \cos \beta \right) - \beta,
\end{eqnarray}
under the condition of $\beta \ge 0$ (See Fig. \ref{collision}).
Here $v$ is the wind velocity, $R_2/a$ is replaced with $f(q)$ 
given in equation(\ref{empirical_roche}), and 
$\eta_{\rm eff}$ is a numerical factor representing the efficiency 
of the energy which is transferred to ablation.  
We assume $\eta_{\rm eff}=1$
in the present calculation but examine the case of $\eta_{\rm eff}=0.3$ 
for comparison.  
By integrating over $\theta$ and using
the wind mass loss rate $\dot M_{\rm w}$ (negative) given by
\begin{equation}
\dot M_{\rm w} = - 4 \pi r^2 \rho v,
\end{equation}
we have
\begin{equation}
{ {G M_2} \over {R_2} }\cdot 
\left( \dot M_2 \right)_{\rm strip} 
=  {1 \over 2} v^2 \dot M_{\rm w} \cdot \eta_{\rm eff} \cdot g(q).
\end{equation}
Here $g(q)$ is the geometrical factor of the red-giant surface 
hit by the wind including dissipation effect 
(Fig. \ref{collision}) and it is only a function of the 
mass ratio $q$. 
Then, the stripping rate of the red-giant envelope is estimated as
\begin{equation}
\dot M_{\rm s} \equiv 
\left( \dot M_2 \right)_{\rm strip} = \eta_{\rm eff} \cdot 
{{v^2 R_2} \over {2 G M_2} } 
\dot M_{\rm w} \cdot g(q),
\label{mass_stripping_rate}
\end{equation}
\begin{equation}
g(q) = \int_{\beta=\pi/2}^{\beta=0} {1 \over 2} \sin^3 \beta
{{R_2^2} \over {r^2}} \sin\theta d\theta, 
\end{equation}
where some numerical values of $g(q)$ are given in Table \ref{tbl-1}.
The total angular momentum loss rate by stripping is also 
estimated as
\begin{equation}
\left( \dot J \right)_{\rm strip} 
= \eta_{\rm eff} \cdot 
{{v^2 R_2} \over {2 G M_2} } 
\dot M_{\rm w} a^2 \Omega_{\rm orb} \cdot h(q),
\end{equation}
\begin{eqnarray}
h(q) &=& \int_{\beta=\pi/2}^{\beta=0} {1 \over 2} \sin^3 \beta
{{R_2^2} \over {r^2}} 
\sin\theta d\theta \cr 
&\times& 
\left({{1} \over {1+q}} - {{R_2} \over {a} } \cos \theta \right)^2,
\end{eqnarray}
\begin{equation}
\left( {\dot J} \over {\dot M} \right)_{\rm strip} 
= \ell_{\rm s} a^2 \Omega_{\rm orb}
= {{h(q)} \over {g(q)}} \cdot a^2 \Omega_{\rm orb},
\end{equation}
where the ablated gas is assumed to have the angular momentum 
at the red-giant surface.
We note that the numerical factor $\ell_{\rm s} = h(q)/g(q)$ 
given in Table \ref{tbl-1} is rather small compared with 
$\ell_{\rm w}$ in equation (\ref{fast_wind_angular_momentum}).
\par
     Including the effect of mass-stripping, we estimate 
the mass transfer rate from the secondary to the primary 
($\dot M_{\rm t} < 0$) as
\begin{equation}
{{\dot M_{\rm t}}\over{M_2}}  =  \left( \left({{\dot R_2}\over
{R_2}}\right)_{\rm EV} - H_3(q) \left({{\dot M_{\rm cr}} \over {M_1}} 
\right) \right)/{H_4(q)}, 
\end{equation}
\begin{equation}
H_3(q)= H_1(q) + {c_1 \over q} H_2(q)
+ 2 \left( \ell_{\rm s} - \ell_{\rm w} \right) (1+q) {c_1 \over q},
\end{equation}
\begin{equation}
H_4(q)= (1 + c_1) H_2(q) 
+ 2 \left( \ell_{\rm s} - \ell_{\rm w} \right) (1+q) c_1, 
\label{hh4q}
\end{equation}
\begin{equation}
c_1 \equiv \eta_{\rm eff} \cdot 
{{v^2 R_2} \over {2 G M_2} } \cdot g(q),
\end{equation}
where $c_1$ is a numerical factor indicating stripping effect as
\begin{equation}
\dot M_{\rm s}= c_1 \dot M_{\rm w},
\end{equation}
and estimated as 
\begin{eqnarray}
c_1 &\sim&
\eta_{\rm eff} \left( {{g(q)} \over {0.025}} \right)
\left({{v} \over {1000 {\rm ~km~s}^{-1}}} \right)^2 \cr 
&\times&\left({{R_2} \over {30 R_\odot}} \right) 
\left({{M_2} \over {2 M_\odot}} \right)^{-1}.
\end{eqnarray}
The stripping effect is important, i.e., $c_1 \sim 1$, when 
the orbital period is longer than $P \sim 30$ d for $\eta_{\rm eff}=1$,
$M_{\rm WD}= 1.0 M_\odot$, $M_2= 2.0 M_\odot$, and $v= 1000$ km s$^{-1}$. 
Function $H_4(q)$ remains negative even for 
$q > 1.15$ because the second term in the right hand 
side of equation (\ref{hh4q}) is always negative, i.e., 
$\ell_{\rm s} < \ell_{\rm w}$. 
Therefore, the mass stripping effect stabilizes the mass transfer 
even for $q > 1.15$, so that the limitation of $q < 1.15$ 
proposed by HKN96 for stable mass transfer is removed.

\placetable{tbl-1}

\section{RESULTS}
      Our progenitor system can be specified by three initial 
parameters: the WD mass $M_{\rm WD,0}$, 
the red-giant mass $M_{\rm RG,0}$, 
and the orbital period $P_0$.
We study three cases P1-P3 (P stands for Progenitor) 
of such close binary evolutions.
We start the calculation when the secondary fills its inner 
critical Roche lobe.  The initial parameters 
$(M_{\rm WD,0}$, $M_{\rm RG,0}$, $P_0$) are summarized 
in Table \ref{tbl_initial_parameters}.  For example, 
$M_{\rm WD,0}= 1.0 M_\odot$, $M_{\rm RG,0}= 2.0 M_\odot$, 
and $P_0= 300$ d for case P1.  
The evolutionary histories are plotted in Figures 
\ref{evolution_wind}-\ref{evolution_unstable}. 
In these three cases, the WDs grow up to $M_{\rm Ia}= 1.38 M_\odot$ 
to trigger an SN Ia explosion as follows.   
\begin{enumerate}
\item[P1)] The mass transfer begins at a rate of
$\dot M_{\rm t} = - 8.7 \times 10^{-7} M_\odot {\rm ~yr}^{-1}$,
and then the WD wind starts to blow at a mass loss rate of 
$\dot M_{\rm w} = - 4.8 \times 10^{-7} M_\odot {\rm ~yr}^{-1}$; and 
the wind then induces the mass stripping at a rate of
$\dot M_{\rm s} = - 2.0 \times 10^{-6} M_\odot {\rm ~yr}^{-1}$.
Thus a half of the transferred matter is blown off in the wind.  
The mass transfer rate gradually decreases because of 
decreasing $M_2/M_{\rm WD}$ but still higher than 
$\dot M_{\rm cr}$ just when the WD mass reaches $M_{\rm Ia}$ 
and explodes as an SN Ia at $t= 7.2 \times 10^5 {\rm ~yr}$ 
during the WIND phase (Fig. \ref{evolution_wind}). 
\item[P2)]
     The mass transfer, the WD wind, and the mass stripping start
as in case P1 at a rate of 
$\dot M_{\rm t}$,
$\dot M_{\rm w}$, and 
$\dot M_{\rm s}$ as summarized in Table \ref{tbl_initial_parameters}.
Thus two thirds of the transferred matter accumulates onto the WD.  
The mass transfer rate is gradually decreasing and becomes lower than
$\dot M_{\rm cr}$ at $t= 6.7 \times 10^5 {\rm ~yr}$. 
The wind stops but nuclear burning is still stable until
the WD mass reaches $M_{\rm Ia}$ to explode as an SN Ia 
at $t= 7.4 \times 10^5 {\rm ~yr}$ (Fig. \ref{evolution_nuc}).  
After the wind stops, 
the progenitor may be observed as a luminous super-soft X-ray source
(SSS).   
\item[P3)]
     The mass transfer, the WD wind, and the mass stripping start
at the rate summarized in Table \ref{tbl_initial_parameters}.
In this case, a large part of the transferred 
matter is accumulating onto the WD.  
The mass transfer rate decreases and becomes lower than
$\dot M_{\rm cr}$ at $t= 5.3 \times 10^5 {\rm ~yr}$. 
The wind stops but the nuclear burning is still stable until
$t= 6.5 \times 10^5 {\rm ~yr}$.  The progenitor may be observed 
as an SSS during this steady hydrogen shell burning phase.
Then nuclear burning becomes unstable to trigger very weak shell
flashes but most of the processed matter 
accumulates onto the WD.  The progenitor may be observed as 
a recurrent nova (RN).  The WD mass eventually reaches 
$M_{\rm Ia}$ to explode as an SN Ia 
at $t= 1.04 \times 10^6 {\rm ~yr}$ (Fig. \ref{evolution_unstable}).  
\end{enumerate}
     We have three cases of the immediate progenitors of our WD+RG 
systems corresponding to cases P1-P3, which are summarized in 
Table \ref{tbl_three_states}, 
i.e., P1) wind phase (denoted by WIND), 
P2) steady nuclear burning phase (denoted by SSS), and P3)
unsteady weak shell flash phase (denoted by RN).
In cases P2 and P3, the SSS phase is rather short 
compared with the wind phase.
Therefore, the frequencies of the SSS phase may be small
in our symbiotic channel to SNe Ia.
  
\placefigure{evolution_wind}
\placefigure{evolution_nuc}
\placefigure{evolution_unstable}
\placefigure{zams10}
\par 
     The final outcome of the binary evolution is summarized 
in the $\log P_0 - M_{\rm d,0}$ plane (Fig. \ref{zams10}).  
The right region of long $P_0$ in the figure represents 
our new results of the WD+RG system. 
In the left region of short $P_0$, 
we also show the results of the WD+MS system 
for comparison (see HKNU99 for details). 
Each grid in the $\log P_0 - M_{\rm d,0}$ plane corresponds to 
the evolutionary model of our wide WD+RG systems
(labeled by ``WD+RG system'')
together with the compact WD+MS systems
(labeled by ``WD+MS system''; HKNU99).  
Here,  $M_{\rm d,0}$
is the mass of the donor, i.e., $M_{\rm RG,0}$ 
(the initial mass of the red-giant component), 
or $M_{\rm MS,0}$ (the initial mass of the slightly 
evolved main-sequence component).
The initial mass of the white dwarf is assumed to be 
$M_{\rm WD,0}= 1.0 M_\odot$.  
\par
     The outcome of the evolution at the end of our calculations 
is classified as follows.
\begin{enumerate}
\item[i)]
  Formation of a common envelope where the mass transfer 
is unstable ($H_4(q) > 0$) at the beginning of mass transfer
(denoted by $\times$). 
\item[ii)]
  SN Ia explosions(denoted by 
$\oplus$ corresponding to case P1, $\bigcirc$ corresponding to case P2, 
and $\odot$ corresponding to case P3), where the WD mass reaches 
1.38 $M_\odot$.
\item[iii)] 
  Novae or strong hydrogen shell flash(denoted by open triangle),
where the mass transfer rate becomes below $\dot M_{\rm low}$.
\item[iv)]
  Helium core flash of the red giant component
(denoted by filled triangle) where
a central helium core flash ignites, i.e., 
the helium core mass of the red-giant reaches 
$0.46 M_\odot$ .   
\end{enumerate}
The region enclosed by the thin solid line produces SNe Ia. 
In HKN96's model, 
this region was limited by $M_{\rm RG,0} < 1.15 M_\odot$ for 
$M_{\rm WD,0} = 1 M_\odot$.  This new area is about ten times or more 
wider than that of HKN96's modeling.
\par
     We also show, in Figure \ref{ztotreg100}, 
other three cases of the initial WD mass,
$M_{\rm WD,0} = 0.8 M_\odot$, $0.9 M_\odot$,
and $ 1.1 M_\odot$ (thin solid) together with  
$M_{\rm WD,0} = 1.0 M_\odot$ (thick solid).  
The regions of $M_{\rm WD,0} = 0.6 M_\odot$ and $0.7 M_\odot$ 
vanish for both the WD+MS and WD+RG systems.
It is clear that the new region of the WD+RG system 
is not limited by the condition of $q < 1.15$, thus being
ten times or more wider than the region of HKN96's model for the other
initial white dwarf masses.
\par
     For $M_{\rm WD,0} > 1.2 M_\odot$, the central density of the WD
reaches $\sim 10^{10}$ g cm$^{-3}$ before heating wave 
from the hydrogen burning layer reaches the
center.  As a result, the WD undergoes collapse due to electron capture
without exploding as an SN Ia (\cite{nom91}).
\par
     To examine the effect of the stripping parameter, $\eta_{\rm eff}$,
we show, in Figure \ref{ztotreg030}, 
the regions of SN Ia explosion for $\eta_{\rm eff}=0.3$.  Here, four 
cases of $M_{\rm WD,0}= 0.9$, $1.0$, $1.1$, 
and $1.2 M_\odot$ are depicted.
The region of $M_{\rm WD,0}= 1.0 M_\odot$ for $\eta_{\rm eff}=0.3$ 
is one third, in area, of the region for $\eta_{\rm eff}=1$ 
(see the region enclosed by the dash-dotted line 
in Fig. \ref{ztotreg100}).
In the limiting case of $\eta_{\rm eff}=0$, we have again 
the constraint of $q< 1.15$ as in HKN96's modeling. 

\placefigure{ztotreg100}
\placefigure{ztotreg030}

\section{DISCUSSION}
\subsection{Recurrent novae as progenitors of SNe Ia}
     First, we introduce a few binary systems which are well understood 
in relation to our SN Ia progenitor model.
T Coronae Borealis (T CrB) and RS Ophiuchi (RS Oph) are 
recurrent novae, which are binaries consisting of a very massive 
white dwarf and a lobe-filling red-giant with orbital periods of 
228 d (\cite{lin88}) and 460 d (\cite{dob94}), respectively.
Two interpretations have been proposed so far on the binary nature 
of T CrB:  one is an episodic mass transfer event (model) 
onto a main-sequence star from a red-giant companion (e.g. \cite{web87}) 
and the other is a thermonuclear runaway event (model) 
on a mass-accreting white dwarf as massive as the Chandrasekhar 
mass limit (e.g., \cite{sel92}).  Ultra-violet lines in its quiescent
phase observed by {\it IUE} strongly indicate the existence of 
a mass-accreting white dwarf instead of a main-sequence star 
(\cite{sel92}). 
\par
     Very rapid decline rates of their light curves indicate 
a very massive white dwarf close to the Chandrasekhar limit, 
that is, $M_{\rm WD} \sim  1.37-1.38 M_\odot$ for T CrB 
(\cite{kat95}, 1999).
It should be noted that Kato (1995, 1999) calculated 
the nova light curves for the white dwarf masses of
1.2, 1.3, 1.35 and 1.377 $M_\odot$, 
where 1.377 $M_\odot$ was chosen as a
limiting mass being just below the mass at the SN Ia explosion in the
W7 model ($1.378 M_\odot$, \cite{nom84}).  Kato found that the
light curve of the $1.377 M_\odot$ model is in better agreement with
observational light curve of T CrB than the lower mass
models.  
\par
     Very recently, other observational supports for a massive white 
dwarf in T CrB have been reported:
one is $M_{\rm WD}= 1.2 \pm 0.2 M_\odot$ by Belczy\'nski
and Mikolajewska (1998) and the other is $M_{\rm WD}= 1.3-2.5 M_\odot$
by Shahbaz et al. (1997).
Belczy\'nski and Mikolajewska derived a
permitted range of binary parameters from amplitude of the ellipsoidal
variability and constraints from the orbital solution of M-giants. The
white dwarf mass is permitted up to $1.44 M_\odot$ under the condition of
a certain mass ratio and inclination of the orbit (in their Fig. 4).  
In Shahbaz et al. (1997), a massive white dwarf of 
$M_{\rm WD}= 1.3-2.5 M_\odot$ in T CrB is also suggested
from the infrared light curve fitting.  Combining these two permitted 
ranges of the white dwarf mass in T CrB, we may conclude that 
a mass of the white dwarf is between $M_{\rm WD}= 1.3-1.4 M_\odot$, 
which is very consistent with the light curve analysis 
($M_{\rm WD} \sim 1.37-1.38 M_\odot$ of T CrB) by Kato (1999).
\par
     However, it is very unlikely that such very massive white 
dwarfs were born at the end of single star evolution 
in a binary (e.g., \cite{wei86}; see also eq.(2) of \cite{yun93}).  
It is more likely that a less massive white dwarf accretes 
hydrogen-rich matter from a red-giant companion and grows up to near 
the Chandrasekhar limit.  If we include the 
mass-stripping effect by the strong WD wind, we easily reproduce 
the present states of T CrB and RS Oph systems. 
\par
     Specifying the initial parameters of 
$M_{\rm WD,0}= 1.0 M_\odot$, $M_{\rm RG,0}= 1.3 M_\odot$, 
and $P_0= 135$ d, we obtain the present state of 
$M_{\rm WD}= 1.37 M_\odot$, $M_{\rm RG}= 0.71 M_\odot$ 
(with a helium core of $M_{\rm He}= 0.35 M_\odot$), 
$P= 228$ d, and the mass transfer rate of 
$\dot M_2 \sim 1 \times 10^{-7} M_\odot$ yr$^{-1}$. 
This set of the initial parameters is shown in Figure 
\ref{zams10} by a star mark ($\star$).  
It seems that T CrB is a critical (failing/succeeding) 
system for SN Ia explosion because this initial model 
corresponds to the lower boundary of the SN Ia region 
for $M_{\rm WD,0}= 1.0 M_\odot$ (Fig. \ref{zams10}).
It should be noted here that a recent analysis by 
Belczy\'nski \& Mikolajewska (1998) shows, contrary to the previous 
results (e.g., \cite{web87}), the mass ratio of T CrB 
to be $q=M_{\rm RG} / M_{\rm WD} \sim 0.6$, which implies a low-mass
binary system and is very consistent with the present 
numerical results. 
\par
     For RS Oph, if we start the calculation with 
$M_{\rm WD,0}= 1.0 M_\odot$, $M_{\rm RG,0}= 1.15 M_\odot$, 
and $P_0= 240$ d, we obtain the present state of 
$M_{\rm WD}= 1.36 M_\odot$, $M_{\rm RG}= 0.60 M_\odot$ 
(with a helium core of $M_{\rm He}= 0.39 M_\odot$), 
$P= 460$ d, and the mass transfer rate of 
$\dot M_2 \sim 1 \times 10^{-7} M_\odot$ yr$^{-1}$. 
RS Oph also seems to be a critical system for SN Ia explosion.
It should be noted that these sets of the initial parameters are
not unique.
\par 
     U Scorpii (U Sco) is also one of the well-known recurrent 
novae.  Light curve fitting indicates a very massive white dwarf 
of $M_{\rm WD}= 1.37-1.38 M_\odot$ (\cite{kat90}, 1995, 1999).
The orbital period is $P= 1.23$ d (\cite{sch95}), thus 
corresponding to the WD+MS system. 
Observations have suggested that the companion of U Sco 
is extremely helium-rich (e.g., \cite{wil81}), 
although its companion is a slightly evolved main-sequence 
star (\cite{sch90}; \cite{joh92}).   
Its evolutionary path has long been regarded as 
a puzzle in the theory of close binary evolution (e.g., \cite{web87}).
Very recently, Hachisu \& Kato (1999b) have elucidated 
the reason why its companion has a helium-rich envelope and,
at the same time, why the white dwarf is so massive as 
the Chandrasekhar mass limit.
\par 
     In Hachisu \& Kato's (1999b) U Sco scenario, 
they start from a progenitor binary system of
$\sim 7-8 M_\odot$ and $\sim 2 M_\odot$ stars
with the initial separation of $\sim 500 ~R_\odot$. 
The primary component has first evolved to fill its Roche lobe 
when the helium core grows to $\sim 1.4-1.6 M_\odot$.  
The binary undergoes a common envelope evolution and shrinks
to the separation of $\sim 10 R_\odot$ between the
naked helium core and the $\sim 2 M_\odot$ 
main-sequence star.  The helium core evolves to fill its Roche lobe 
and stably transfers almost pure helium 
onto the secondary because of the mass ratio $M_1/M_2 \lesssim 0.79$.  
As a result, the secondary becomes a helium-rich 
star.  
\par
     After the helium envelope of the primary is exhausted, 
the primary becomes a carbon-oxygen (C+O) white dwarf 
of $0.9-1.0 M_\odot$ and 
the secondary grows in mass to $\sim 2.5 M_\odot$.
When the secondary slightly evolves to fill its Roche lobe,
it transfers helium-rich matter onto the C+O white dwarf 
on a thermal time scale.
The white dwarf burns hydrogen atop the surface at a critical rate 
of $\dot M_{\rm cr} \sim 2.0 \times 10^{-6} 
(M_{\rm WD}/M_\odot - 0.40) ~M_\odot$ yr$^{-1}$ 
for helium-rich matter and blows excess matter in winds.
The white dwarf now grows to near the Chandrasekhar mass limit
and the mass transfer rate decreases to a few to several 
times $10^{-7} M_\odot$ yr$^{-1}$.  
These pictures seems to be very consistent 
with the present distinct observational aspects of U Sco.
\par
     To summarize, the recurrent novae seem to be a critical system 
for SN Ia explosion.  
Recurrent novae are morphologically divided into three groups; 
dwarf companion, slightly evolved main-sequence companion, 
and red-giant companion (e.g., \cite{sch95}).
T CrB ($P_{\rm orb}= 228$ d) and RS Oph ($P_{\rm orb}= 460$ d)
belong to the last group of red-giant companion.
U Sco ($P_{\rm orb}= 1.23$ d; \cite{sch95}) 
and V394 CrA ($P_{\rm orb}= 0.758$ d; \cite{sch90}) 
belong to the middle group of the slightly evolved main-sequence 
companions.  
Two of the three subgroups of the recurrent novae
correspond to our progenitors (WD+MS/WD+RG systems).
This close relation between the recurrent novae and our progenitors
strongly support our scenario of SN Ia progenitors.

\subsection{Yungelson \& Livio's criticism}
     Based on the population synthesis analysis, 
Yungelson \& Livio (1998) claimed that almost no realization 
frequency is derived for the original HKN96's WD+RG model. 
First, we briefly explain their analysis why the original 
model by HKN96 does not produce enough number of SNe Ia.  
Second, we point out that 
very wide binaries with the initial separation 
of $a_i \gtrsim 1500 ~R_\odot$, which were not included in 
Yungelson \& Livio's (1998) analysis, are essentially important in our 
SN Ia modeling.
\par
     A more massive component (mass of $M_{1,i}$) of a binary 
first evolves to 
a red-giant (AGB stage) and fills its inner critical Roche lobe. 
After a common envelope phase, the more massive component leaves 
a C+O WD and the separation of the binary decreases by a factor of 
\begin{equation}
{{a_f} \over {a_i}} \sim \alpha_{\rm CE} 
\left({{M_{\rm WD}} \over {M_{1,i}}} \right) 
\left({{M_2} \over {M_{1,i}-M_{\rm WD}}} \right),
\end{equation}
where $\alpha_{\rm CE}$ is the efficiency factor of 
common envelope evolutions, $a_f$ ($a_i$) 
the final (initial) separation, and 
$M_2$ the mass of the secondary. 
Adopting a standard value of $\alpha_{\rm CE}=1$, 
we obtain $a_f/a_i  \sim 1/40 - 1/50$ 
for $M_{\rm WD} \sim 1 ~M_\odot$ 
and $M_2 \sim 1 ~M_\odot$, because a $\sim 1 ~M_\odot$ WD descends 
from a main-sequence star of 
$M_{1,i} \sim 7-8 ~M_\odot$ (e.g., \cite{wei86}; 
\cite{yun95}).  Yungelson \& Livio (1998) assumed that 
the separation of interacting binaries is {\it smaller than} 
$a_i \lesssim 1500 ~R_\odot$. 
Then, the most wide binaries has the separation of 
$a_f \lesssim 30-40 ~R_\odot$  after the common envelope evolution.  
Its orbital period is $P_0 \lesssim 20$ d for $M_{\rm WD,0} \sim 1 ~M_\odot$ 
and $M_{\rm RG,0} \sim 1 ~M_\odot$.  
There is no SN Ia region of the WD+RG systems 
for $P_0 \lesssim 20$ d as seen from Figure \ref{zams10}. 
Thus, they concluded that we cannot expect any SN Ia explosions 
from the right SN Ia region (WD+RG system) of Figure \ref{zams10}.
\par
     If the WD+RG evolution starts only from an initial condition 
of $P_0 \lesssim 20$ d and $M_{\rm WD,0} \sim 1 M_\odot$, 
however, the present states of T CrB or RS Oph cannot be 
reached because of low mass transfer rates 
of $|\dot M_{\rm t}| \lesssim 1 \times 10^{-7} M_\odot$ yr$^{-1}$ 
(see also Figs. \ref{zams10} and \ref{ztotreg100}).  
Thus, the existence of recurrent 
novae T CrB and RS Oph seems to be 
against the above Yungelson \& Livio's conjecture.  The reason why 
T CrB or RS Oph are failed to be reproduced 
in Yungelson \& Livio's modeling is due to  
their assumption of $a_i \lesssim 1500 ~R_\odot$. 
In what follows, we show that such wide WD+RG binaries as
$P_0 \sim 100-1000$ d are born from initially very wide 
binaries with $a_i \sim 1500-40000 ~R_\odot$.
\par
     A star with the zero-age main-sequence mass of 
$M_{1,i} \lesssim 8 ~M_\odot$ ends up 
its life by ejecting its envelope in a
wind of relatively slow velocities ($v \sim 10-40$ km s$^{-1}$).  
These wind velocity is as low as the orbital velocity of 
binaries with the separation of $a_i \sim 1500-40000 ~R_\odot$
for $M_{1,i} \sim 7 ~M_\odot$ and $M_{2,i} \sim 1 ~M_\odot$.
When the wind velocity is as low as the orbital velocity, 
the numerical factor $\ell_{\rm w}$ 
in equation (\ref{specific_angular_momentum_wind}) increases to 
\begin{equation}
\ell_{\rm w} \approx 1.7 - 0.55
\left( {{v} \over {a \Omega_{\rm orb}}} \right)^2,
\label{slow_wind_angular_momentum}
\end{equation} 
mainly because outflowing matter can get angular momentum 
from the binary motion by torque during its journey 
(see Appendix A). 
Here, $v$ is the radial component of the wind velocity 
near the inner critical Roche lobe and the limiting case of
$\ell_{\rm w}=1.7$ for $v=0$ was obtained by
Nariai 
(1975) and
Nariai \& Sugimoto 
(1976) for a test particle simulation ejecting from the outer
Lagrangian points and Sawada, Hachisu, \& Matsuda 
(1984) for a 2-D (equatorial plane) hydrodynamical simulation blowing 
a very slow wind from the primary surface which fills
the inner critical Roche lobe.  
\par
     Combining the two expressions, we obtain
\begin{equation}
\ell_{\rm w} \approx \max\left[ 1.7 - 0.55
\left( {{v} \over {a \Omega_{\rm orb}}} \right)^2,
\left( {{q} \over {1+q}} \right)^2 \right],
\label{both_wind_angular_momentum}
\end{equation} 
which is a good approximation to $\ell_{\rm w}$ in the region of 
the wind velocity from zero to infinity.  Switching from 
equation (\ref{fast_wind_angular_momentum}) 
to equation (\ref{slow_wind_angular_momentum}) occurs at 
$v \sim 1.5 a \Omega_{\rm orb}$ for $q=2$.
\par 
     If we assume that slow winds blow from the primary and 
the mass of the secondary does not change ($\dot M_2=0$), 
the separation is calculated from  
\begin{eqnarray}
{{\dot a} \over a} &=& {{\dot M_1 + \dot M_2} \over {M_1+M_2}} 
- 2 {{\dot M_1} \over {M_1}} - 2 {{\dot M_2} \over {M_2}} 
+ 2 {{\dot J} \over {J}} \cr
&=& \left( 2 \ell_{\rm w} {{M_1+M_2} \over {M_2}} 
+ {{M_1} \over {M_1+M_2}} -2 \right)
{{\dot M_1} \over {M_1}} \cr
& \approx & \left( 2 \ell_{\rm w} {{M_1} \over {M_2}} -1 \right) 
{{\dot M_1} \over {M_1}}, 
\qquad (M_1 \gg M_2).
\label{separation_decrease_by_wind}
\end{eqnarray}
Therefore, the critical value of $\ell_{\rm w}$ is 
about $0.5 q = 0.5 (M_2/M_1)$ for shrinking/expanding of the separation. 
When $v \sim a \Omega_{\rm orb}$, the systemic loss 
of the angular momentum is estimated as 
$\ell_{\rm w} \approx 1.7 - 0.55(v/a \Omega_{\rm orb})^2 \sim 1.0$ 
from equation (\ref{both_wind_angular_momentum}).
We therefore have $\dot a/a \approx 2 \dot M_1 / M_2 < 0$.
\par
     Once the binary system begins to shrink, its evolution becomes 
similar to a common envelope evolution 
(see {\it stages B-D} in Fig. \ref{channel}). 
As the separation shrinks, the orbital velocity 
of $a \Omega_{\rm orb}$ increases.  If the wind velocity is
almost constant, the ratio of $v/a \Omega_{\rm orb}$ 
in equation (\ref{both_wind_angular_momentum}) becomes smaller 
and smaller and the shrinking is accelerated more and more.
Thus, the separation is reduced by a factor of $1/40-1/50$, i.e., 
$a_f \sim 30-1000 ~R_\odot$ 
for $M_{1,i} \sim 7 ~M_\odot$ and $M_{2,i} \sim 1 ~M_\odot$.
The orbital period becomes $P_0 \sim 15-3000$ d 
for $M_{\rm WD,0} \sim 1 ~M_\odot$ and $M_2 \sim 1 ~M_\odot$
({\it stage D} in Fig. \ref{channel}).
These initial sets of the parameters are very consistent with 
the initial conditions of our WD+RG progenitor systems. 
\par
     To summarize, we must include binaries with the separation of 
\begin{equation}
a_i \lesssim 5500 R_\odot \left( {{\xi} \over {1.7}} \right)^2
\left( {{M_{1,i}+M_{2,i}} \over {M_\odot}}
\right) \left({{10 \mbox{~km~s}^{-1}} \over {v}}\right)^2,
\label{binary_contraction}
\end{equation}
into the category of {\it interacting binaries}, where
$v$ is the velocity of slow wind (super wind) at the end of 
AGB evolution and $\xi \sim 1.5-1.7$ is a numerical factor 
defined by
\begin{equation}
\xi^2 \equiv 3.0 - 0.9 {{q(1+2q)} \over {(1+q)^2}}.
\end{equation}
Here, we take the critical wind velocity 
of shrinking/expanding as 
\begin{equation}
v_{\rm cr} = \xi a \Omega_{\rm orb},
\end{equation}
at the beginning of super wind phase. 

\subsection{SN Ia frequency}
     We estimate the SN Ia rate in our Galaxy coming from our 
WD+RG/WD+MS systems by using equation (1) 
of Iben \& Tutukov (1984), i.e.,
\begin{equation}
\nu = 0.2 \cdot \Delta q \cdot 
\int_{M_A}^{M_B} {{d M} \over M^{2.5}} \cdot \Delta \log A 
\quad {\rm yr}^{-1},
\label{realization_frequency}
\end{equation}
where $\Delta q$, $\Delta \log A$, $M_A$, and $M_B$ are 
the appropriate ranges of the mass ratio, of the initial separation, 
and the lower and the upper limits of the primary mass
for SN Ia explosions in solar mass units, respectively. 
The estimated rate of the WD+RG/WD+MS systems is close 
to the observed rate in our Galaxy, $\nu \sim 0.003$ yr$^{-1}$ 
(e.g., \cite{cap97}; see also \cite{yun98}), as will be shown below.  

\subsection{WD+RG systems}
     For the WD+RG progenitors, 
we assume that the initial region of the separation includes 
$a_i \sim 1500 - 40000 ~R_\odot$ as well as 
$a_i \lesssim 1500 ~R_\odot$ (see discussions above in \S 4.2). 
Dividing the initial white dwarf mass of $M_{\rm WD,0}$ 
into four intervals, i.e., 
$0.8-0.9 M_\odot$, 
$0.9-1.0 M_\odot$, 
$1.0-1.1 M_\odot$, and 
$1.1-1.2 M_\odot$,
we estimate the realization frequencies.
The mass range of $M_{{\rm WD,0}} < 0.7 M_\odot$ is not included
because no SN Ia explosions are expected for such low initial mass WDs.
We omit the range of $M_{{\rm WD,0}}=0.7-0.8 M_\odot$ because 
its realization frequency is too small to contribute to the SN Ia rate
as seen in Figures \ref{ztotreg100} and \ref{ztotreg030}. 
To estimate the initial separation(or orbital period), $\log A$, and 
the initial lower/upper masses, $M_A$ and $M_B$, of our 
WD+RG systems, we need to obtain the zero-age main sequence mass 
of the primary component ($M_{1,i}$) and the contraction factor 
after the first common envelope phase.  In single star evolutions,
$0.7-1.2 M_\odot$ white dwarfs descend from 
stars with the zero-age main sequence mass of 
$M_i \sim 3-8 M_\odot$, 
i.e., $M_A \sim 3 M_\odot$ and $M_B \sim 8 M_\odot$. 
More precisely, using Yungelson et al.'s (1995) equation (11) gives  
the final core mass (C+O WD mass) 
vs. the zero-age main sequence mass relation, 
\begin{equation}
\log {{M_{\rm C+O}} \over {M_\odot}}= 
-0.22 + 0.36 \left( \log {{M_0} \over {M_\odot}} \right)^{2.5},
\label{mass_relation_CO}
\end{equation}
as numerically summarized in Table \ref{tbl_contraction_factor}, 
where $M_{\rm WD,0} = M_{\rm C+O}$.
The initial separation should be larger than 
\begin{equation}
a_i > {{R_1 ({\rm AGB})} \over {f(M_{1,i}/M_{2,i})}} 
\approx 2 R_1 ({\rm AGB}),
\label{minimum_separation_CO}
\end{equation}
in order that the C+O core grows up to $M_{\rm C+O}= M_{\rm WD,0}$. 
Here $f(q) \approx 0.5$ for $q \equiv M_{1,i}/M_{2,i} \sim 2-7$ 
and the radius of stars at the AGB phase is given by
\begin{equation}
{{R_1 ({\rm AGB})} \over {R_\odot}} =
1050 \left( {{M_{\rm C+O}} \over {M_\odot}} - 0.5 \right)^{0.68},
\label{radius_AGB_CO}
\end{equation}
(\cite{ibe84}). 
For example, the initial separation should be larger than 
$a_i \sim 1,200 ~R_\odot$  for $M_{\rm WD,0}=1.0 M_\odot$ 
as summarized in Table \ref{tbl_contraction_factor}.
\par 
     For the binary of $M_{\rm WD,0}=1.0 M_\odot$ 
and $M_{2,i} = 1 M_\odot$, the contraction factor is estimated 
to be $1/37$ by assuming 
the common envelope efficiency factor of $\alpha_{\rm CE}=1$. 
The range of the separation after common envelope evolution becomes 
$a_f \sim 32-1,120  ~R_\odot$ (corresponding to 
$P_0 \sim 15-3,070$ d) 
because of $a_i \sim 1,200-41,700 ~R_\odot$ as summarized 
in Table \ref{tbl_contraction_factor}. 
The orbital period of $P_0 \sim 15-3,070$ d covers 
the SN Ia region (WD+RG system) of Figure \ref{zams10}. 
The binary parameters for other $M_{\rm WD,0}$ are summarized in
Table \ref{tbl_contraction_factor}.  
The regions of the orbital period, $\log P_0$, 
also covers the SN Ia region (WD+RG system) of Figure \ref{ztotreg100}.
Here we assume $v= 10$ km s$^{-1}$ 
and $v < v_{\rm cr} = \xi a\Omega_{\rm orb}$ to meet 
the condition for the binary to contract (\S 4.2).  
Since the region of $\log A$ ($\log a_i$) is shifted in parallel to 
the region of $\log a_f$ by the contraction factor,
the probability frequency for $\Delta \log A$ is the same as for
$\Delta \log a_f$.
Then we approximately set as 
\begin{equation}
\Delta \log A \approx \Delta \log P_0 \cdot {2 \over 3}, 
\label{separation_period_2/3}
\end{equation}
where $\Delta \log P_0$ is taken from the SN Ia region in 
Figure \ref{ztotreg100} and 
the factor of $2/3$ comes from 
the conversion between the period and the separation. 
Substituting $\Delta \log A = 0.6 \cdot 2/3$, 
$\Delta q = 2.6/4.48-1.2/5.60=0.37$, 
$M_A = 4.48$, $M_B = 5.60$, we obtain 
$\nu_{{\rm WD},0.8-0.9} = 0.0006$. 
\par
     The SNe Ia rates for other WD mass intervals are summarized 
in Table \ref{tbl_realizaton_frequency}.  
Then, the summation of SN Ia rates for 
three intervals ($0.8-0.9 M_\odot$, $0.9-1.0 M_\odot$, and 
$1.0-1.1 M_\odot$) gives $\nu_{\rm RG}= 0.0017$ yr$^{-1}$, 
which is large enough to explain the dominant part of the SN Ia rate 
in our Galaxy.  If we further include the WD mass range of 
$M_{\rm WD,0}=1.1-1.2 M_\odot$, which is not shown 
in Table \ref{tbl_realizaton_frequency},
the realization frequency increases to $\nu_{\rm RG}= 0.0022$ yr$^{-1}$.
Here, we have not shown the region for $M_{\rm WD,0}=1.2 M_\odot$
in Figure \ref{ztotreg100} 
because Webbink et al.'s (1983) empirical formula is valid for 
$M_{2,0} < 2.5-3.0 M_\odot$ with a degenerate helium core.  
The range of $M_{\rm WD,0}=1.2 M_\odot$ exceeds this limit.
\par
     To examine the effect of the stripping parameter, $\eta_{\rm eff}$,
we estimate the realization frequency for $\eta_{\rm eff}=0.3$ 
as summarized in Table \ref{tbl_realization_0.33}.  
The parameter region shrinks to one third 
in area compared with the case of $\eta_{\rm eff}=1$, so that 
the realization frequency is reduced to about one third 
of $\eta_{\rm eff}=1$ case, i.e., $\nu_{\rm RG}= 0.0008$ yr$^{-1}$.

\subsection{WD+MS systems}
     For the WD+MS progenitors, HKNU98 have found 
a new evolutionary path, which has 
not been taken into account in the previous works 
(e.g., \cite{rap94}; \cite{dis94}; \cite{yun96}; \cite{yun98}), 
and estimated the realization frequency to be as large as 
$\nu_{\rm MS}= 0.001$ yr$^{-1}$.  We briefly follow their 
new evolutionary path in the following and discuss the total 
rate of SN Ia explosions:
If a $\sim 1 M_\odot$ C+O WD is descending from an AGB star,
its zero-age main sequence mass is $\sim 7 M_\odot$ by 
equation (\ref{mass_relation_CO}) and the binary separation 
is larger than $a_i \sim 1350 ~R_\odot$ if the secondary mass 
is $\sim 2 M_\odot$.
Its separation shrinks to $a_f \sim 70 ~R_\odot$ 
after the common envelope evolution with 
$\alpha_{\rm CE} = 1$ (the secondary mass of $\sim 2 M_\odot$).
Then the orbital period becomes $P_0 \sim 40$ d and too long 
to become an SN Ia as seen in Figure \ref{zams10}.  
Therefore, the C+O WD comes not from an AGB star 
having the radius of equation (\ref{radius_AGB_CO})
but from a helium star whose
hydrogen-rich envelope has been stripped away in the first common 
envelope evolution (at the red-giant phase with a helium core).
Then, we follow the evolution of a binary consisting of a helium star 
and a main-sequence star.  
\par
     To estimate the decrease in the separation 
after the common envelope phase, we use the radius to helium core mass 
($R_1 - M_{\rm 1,He}$) relation, 
which are taken from tables given by Bressan et al. (1993). 
As an example, let us consider a pair of $7 M_\odot + 2.5 M_\odot$
with the initial separation of $a_i \sim 50-600 R_\odot$.
The binary evolves to SN Ia through the following stages:
\begin{enumerate}
\item[1)]  
When the mass of the helium core grows to 
$1.0 M_\odot < M_{\rm 1, He} < 1.4 M_\odot$,
the primary fills its Roche lobe and the binary undergoes 
a common envelope evolution.   
\item[2)] 
After the common envelope evolution,
the system consists of a helium star and a 
main-sequence star with a relatively compact separation of
$a_f \sim 3-40 ~R_\odot$ and $P_{\rm orb} \sim 0.4 - 20$ d.
\item[3)]
The helium star contracts and ignites
central helium burning to become a helium main-sequence star.  
The primary stays at the helium main-sequence 
for $\sim 1 \times 10^7$ yr (e.g., \cite{pac71}).
\item[4)]
After helium exhaustion, a carbon-oxygen core develops.
When the core mass reaches $0.9-1.0 M_\odot$,
the helium star evolves to a red-giant and fills again its 
inner critical Roche lobe.  Almost pure helium is
transferred to the secondary because the mass transfer is 
stable for the mass ratio $q=M_1/M_2 < 0.79$.
The mass transfer rate is as large as 
$\sim 1 \times 10^{-5} M_\odot$ yr$^{-1}$ but the mass-receiving 
main-sequence star ($\sim 2-3 M_\odot$) does not expand for 
such a low rate (e.g., \cite{kip77}).
\item[5)]
The secondary has 
received $0.1-0.4 M_\odot$ (almost) pure helium and, 
as a result, it becomes a helium-rich star as observed in
U Sco (e.g.,\cite{wil81}; \cite{bar81}; \cite{han85}; 
\cite{sek88}).  The separation and thus the orbital period 
gradually increases during the mass transfer phase.  
The final orbital period becomes $P_{\rm orb} \sim 0.5-40$ d.  
\item[6)]
An SN Ia explosion occurs when $P_0 = 0.4 - 5$ d 
and $M_2= M_{\rm MS,0} \sim 2-3 M_\odot$ in 
the $\log P_0$-$M_{\rm d,0}$ plane
for the system of $M_1= M_{\rm WD,0} \sim 0.9-1.0 M_\odot$ 
as seen in Figure \ref{ztotreg100}.  
\end{enumerate}
Therefore, the above pair of $7 M_\odot + 2.5 M_\odot$ 
can be a progenitor of SNe Ia if the initial separation 
is between $50-150 R_\odot$, which initiates a common 
envelope evolution at the helium core mass of 
$M_{\rm 1,He}= 1.0-1.2 M_\odot$, corresponding to 
the initial orbital period of $P_{\rm orb,0}= 0.5-5$ d for
the WD+MS systems in Figure \ref{zams10}.
In this case, we have $\Delta \log A= \log 150 - \log 50= 0.5$
and $q= 2.5/7= 0.36$ in equation (\ref{realization_frequency}).
\par
     Calculating 25 pairs of $M_{1,i}$ and $M_{2,i}$, 
HKNU99 have obtained the parameter region of SN Ia explosion as
$\Delta q= 0.4$, $\Delta \log A= 0.5$, $M_A= 5.5$, and $M_B= 8.5$.
Substituting these values into equation (\ref{realization_frequency}), 
we obtain the SN Ia rate of $\nu_{\rm MS}= 0.001$ yr$^{-1}$.  
Thus HKNU99 have shown that 
the frequency of the WD+MS systems is about one third
of the inferred rate in our Galaxy, which is much larger than that 
of Yungelson \& Livio's (1998) estimation.  
It should be noted that Yungelson \& Livio (1998) have obtained 
the birth rate of $1 \times 10^{-3}$ yr$^{-1}$ for their 
models 15 and 16 by relaxing all the constraints on the mass ratio of 
their binary models, although it is not a realistic case.
\par
     The orbital velocity of the WD+MS systems is much faster than that 
for the WD+RG systems, i.e., $a \Omega_{\rm orb} \sim 400$ km s$^{-1}$
for $M_{\rm WD}= 1.0 M_\odot$ and $M_{\rm MS}= 2.0 M_\odot$ at 
the zero-age main sequence.  Then, the switching from 
equation (\ref{fast_wind_angular_momentum}) 
to equation (\ref{slow_wind_angular_momentum}) occurs at 
$v \sim 1.5 a \Omega_{\rm orb} \sim 600$ km s$^{-1}$.
This means that the wind velocity has to be faster than 
$\sim 600$ km s$^{-1}$ in order to avoid the formation of 
a common envelope. Otherwise, winds carry large specific angular 
momentum and drastically shorten the separation to enhance 
the mass transfer and to eventually form a common envelope.
Our strong winds satisfy the condition of $v \gtrsim 600$ km s$^{-1}$ 
and this supports Li \& van den Heuvel's evolutionary process 
and also HKNU99's modeling on the WD+MS systems.
\par
     To summarize, the contribution of the WD+MS systems to the SN Ia 
rate in our Galaxy is about one third of the inferred rate in our Galaxy.
The total SN Ia rate of the WD+MS/WD+RG systems 
becomes $\nu_{\rm RG+MS}=\nu_{\rm RG}+\nu_{\rm MS}= 0.003$ yr$^{-1}$, 
which is close enough to the inferred rate of our Galaxy.

\def\mdot{$\dot M$}
\def\msy{$M_\odot$ yr$^{-1}$}
\def\kms{km s$^{-1}$}
\def\e#1{$\times$ $10^{#1}$ }
\def\ee#1{$10^{#1}$ }

\subsection{Observational detection of hydrogen}
In our scenario, the WD winds form a circumstellar envelope 
around the binary systems prior to the explosion, which may emit X-rays,  
radio, and H$\alpha$ lines by shock heating 
when the ejecta collide with the circumstellar envelope. 
In HKN96's model of binary evolution, the mass accretion rate
decreases well below 1 \e{-6} \msy~ as the white dwarf mass gets close
to the Chandrasekhar limit.  Thus the strong wind has ceased when the
white dwarf explodes (HKN96).  In contrast, the mass accretion rate in
the present models is still as high as 1 \e{-6} \msy~ for some of the
white dwarfs near the Chandrasekhar limit, so that such 
a white dwarf explode in the strong wind phase.
\par
     Our strong wind model of case P1 predicts the
presence of circumstellar matter around the exploding white dwarf.
Whether such a circumstellar matter is observable depends on its
density.  The wind mass loss rate from the white dwarf near the
Chandrasekhar limit is 
as high as \mdot~ $\sim$ 1 \e{-8}$-$1 \e{-7} \msy~
and the wind velocity is $v=$ 1000 \kms~ (Fig. \ref{dmdtvelx70z02}).
Despite the relatively high mass loss rate, the circumstellar density
is not so high because of the high wind velocity.  For steady wind,
the density is expressed by $\dot M /v$ ($=4 \pi r^2 \rho$).  
Normalized by the typical
red-giant wind velocity of 10 \kms, the density measure of our white
dwarf wind is given as \mdot$/v_{10} \sim 1$ \e{-10}$-1 $\e{-9} 
\msy, where $v_{10} = v$/10 \kms.

Behind the red-giant, matter stripped from the red-giant component 
forms a much dense circumstellar tail.  
Its rate is as large as $\sim 1 \times 10^{-7} M_\odot$ yr$^{-1}$
with the velocity of $\sim 100$ km s$^{-1}$.  The density measure 
of the dense red-giant wind thus formed is 
\mdot/$v_{10} \sim$ 1 \e{-8} \msy.

Further out, circumstellar matter is produced from the wind from the
red-giant companion, which is too far away to cause significant
circumstellar interaction.

For cases P2 and P3, winds from the WD have stopped before the
explosion.  Therefore, circumstellar matter is dominated by the wind
from the red-giant companion whose velocity is as low as $\sim 10$ km
s$^{-1}$.

At SN Ia explosion, ejecta
would collide with the circumstellar matter,
which produces shock waves propagating both outward and inward.  At
the shock front, particle accelerations take place to cause radio
emissions.  Hot plasmas in the shocked materials emit thermal X-rays.
The circumstellar matter ahead of the shock is ionized by X-rays and
produce recombination H$\alpha$ emissions (\cite{cum96}).
Such an interaction has been observed in Type Ib, Ic, and II
supernovae, most typically in SN 1993J (e.g., \cite{suz95}
and references therein).

For SNe Ia, several attempts have been made to detect 
the above signature of circumstellar matter.  
There has been no radio and X-ray detections so far.  
The upper limit set by X-ray observations of SN
1992A is \mdot$/v_{10} = (2-3)$ \e{-6} \msy~ (\cite{sch93}).
Radio observations of SN 1986G have provided the most strict upper
limit to the circumstellar density as \mdot/$v_{10} =$ 1 \e{-7} \msy~
(\cite{eck95}).  This is still $10-100$ times higher than the
density predicted for the white dwarf wind for case P1.
For cases P2 and P3, if the wind mass loss rate from the red-giant is
significantly higher than 1 \e{-7} \msy, radio detection could be
possible for very nearby SNe Ia as close as SN 1986G.  (Note also that
SN 1986G is not a typical SN Ia but a subluminous SN Ia.)

For H$\alpha$ emissions, Branch et al. (1983) noted a small, narrow
emission feature at the rest wavelength of H$\alpha$, which is
blueshifted by 1800 \kms~ from the local interstellar Ca II
absorption.  Though this feature was not observed 5 days later, such
high velocity hydrogen is expected from the white dwarf wind model.  For SN
1994D, Cumming et al. (1996) obtained the upper limit of \mdot/$v_{10}
=$ 6 \e{-6} \msy.  Further attempts to detect H$\alpha$ emissions are
highly encouraged (\cite{lun97}).

\section{CONCLUSIONS}
     Progenitors of SNe Ia have not been identified yet 
either theoretically or observationally. 
In the present paper, 
we propose a new evolutionary process which produces a wide enough
symbiotic channel to SNe Ia.  In this channel, the 
white dwarf accreting mass 
from the lobe-filling evolved companion (red-giant) 
grows to the Chandrasekhar limit and explodes as an SN Ia.
In what follows we summarize our main results: 

\begin{enumerate}
\item In HKN96, we have found the most 
important mechanism in our symbiotic channel to SNe Ia:
when the mass accretion rate onto the white dwarf exceeds 
a critical rate of $\dot M_{\rm cr} \approx 0.75 \times 10^{-6} 
(M_{\rm WD}/M_\odot - 0.40) M_\odot {\rm ~yr}^{-1}$, 
a strong optically thick wind blows from the white dwarf. 
It stabilizes the mass transfer even if the red-giant has 
a deep convective envelope.  
HKN96 have shown that the mass transfer becomes
stable only when the mass ratio of RG/WD is smaller than 1.15.  
In the present study, we have found that 
this constraint is removed when the effect of 
mass-stripping from the red-giant envelope by the strong wind 
is large enough.
Thus the symbiotic channel produces SNe Ia
for a much (more than ten times) wider 
range of the binary parameters than that of 
HKN96's estimation, thus being able to 
account for the dominant part 
of the inferred rate of SNe Ia in our Galaxy. 
It should be noticed, however, that the realization frequency depends 
on the efficiency factor of mass-stripping $\eta_{\rm eff}$ and 
the realization frequency is consistent with the inferred rate
in our Galaxy for $\eta_{\rm eff} \sim 0.3-1.0$.

\item Yungelson \& Livio (1998) estimated the realization frequency 
of HKN96's original model and concluded 
that almost no realization 
frequency is derived because the binary shrinks too much 
($P_{\rm orb} < 20$ d) to produce an SN Ia after the first 
common envelope evolution.  In their population synthesis code, however,  
Yungelson \& Livio (1998) assumed that the initial separation is 
{\it smaller than} $\sim 1500 ~R_\odot$, 
i.e., $a_i \lesssim 1500 ~R_\odot$.  This assumption neglects 
the effect of slow winds at the end of stellar evolution. 
When the velocity of the slow wind is as small as 
the orbital velocity, the slow wind gets angular momentum 
through torque by the binary motion and, as a result, 
the binary shrinks so much.  When the wind velocity is as slow as 
$\sim 10$ km s$^{-1}$, we must include the binaries with the separation
of $a_i \sim 1,500-40,000 ~R_\odot$.  Such an extremely wide binary 
will shrink to an appropriate range of the separation 
($20$ d $< P_0 < 3000$ d for a pair of $M_{\rm WD,0} \sim 1 M_\odot$ and 
$M_{2,i} \sim 1 M_\odot$) after the slow wind (super wind) 
and the common envelope evolution. 
If this effect is included 
in the calculation of SN Ia rate, we have a reasonable value 
of the realization frequency, $\nu_{\rm RG}= 0.002$ yr$^{-1}$, 
for our white dwarf plus red-giant (WD+RG) systems.

\item Yungelson \& Livio (1998) estimated 
the realization frequency of 
Li \& van den Heuvel's (1997) white dwarf plus main sequence star 
(WD+MS) model and concluded that the total frequency 
of their modified HKN96's model 
and Li \& van den Heuvel's model
does not exceed 0.0002 yr$^{-1}$, a tenth of the inferred rate.
However, we believe that an important evolutionary path was not 
included in Yungelson \& Livio's (1998) analysis, that is, 
the primary's helium star phase: the primary becomes 
a naked helium star after the first common envelope phase 
if the mass transfer begins at the phase of a red-giant  
with a helium core.
This helium star eventually leaves a C+O WD by transferring
the helium envelope to the secondary.
As a result, the secondary may have an extremely helium-rich 
envelope such as in U Sco 
(e.g., \cite{wil81}; \cite{bar81}; \cite{han85}; 
\cite{sek88}).  This evolutionary path indicates that 
the secondary of the WD+MS systems has a helium-enriched envelope 
thus forming an accretion disk which shows strong helium lines as seen in
the luminous supersoft X-ray sources (e.g., \cite{kah97} 
for a recent review).  Including this evolutionary path 
in the estimation of SN Ia rate, HKNU99 
have obtained a much larger frequency of 
$\nu_{\rm MS}= 0.001$ yr$^{-1}$ for the WD+MS systems than that 
estimated by Yungelson \& Livio (1998).  Thus the total frequency 
of our WD+RG/WD+MS systems is as large as 
$\nu= \nu_{\rm RG} + \nu_{\rm MS}= 0.003$ yr$^{-1}$, 
which is consistent with the inferred rate in our Galaxy.

\item Recurrent novae are morphologically divided into three groups; 
dwarf companion, slightly evolved main-sequence companion, 
and red-giant companion (e.g., \cite{sch95}).
T CrB ($P_{\rm orb}= 228$ d) and RS Oph ($P_{\rm orb}= 460$ d)
belong to the last group of red-giant companion.
It has been argued that their white dwarf masses are very close to 
the Chandrasekhar mass limit.   
These systems correspond to near the lower boarder of our SN Ia region
(WD+RG) as shown in Figure \ref{zams10} and its evolutionary path 
is reasonably understood by our WD+RG systems.  On the other hand, 
U Sco ($P_{\rm orb}= 1.23$ d) and V394 CrA ($P_{\rm orb}= 0.758$ d) 
belong to the middle group of slightly evolved main-sequence 
companion.  
Our WD+MS model naturally yields a helium-enriched envelope 
of the secondary star as well as a near Chandrasekhar mass 
limit white dwarf as has been observationally suggested.  
The evolutionary path of the WD+MS systems 
is also very consistent with the middle group of recurrent novae.

\item The immediate progenitors in our symbiotic channel to SNe Ia may
be observed as a symbiotic star.  The photospheric temperature of the
mass-accreting white dwarf is kept around $T_{\rm ph} \sim 1 \times
10^5 - 2 \times 10^5$ K during the wind phase.  The hot star may not
be observed in X-rays during the strong wind phase due to
self-absorption by the wind itself.  Some progenitors stop blowing
the wind before the SN Ia explosion, thus being observed as a luminous
supersoft X-ray source.  In some progenitors, very weak hydrogen shell
flashes are triggered before the SN Ia explosion; such a progenitor
may be observed as a recurrent nova like T CrB or RS Oph.

\item Radio emission from the circumstellar gas is predicted.  If the
white dwarfs explode in the strong wind phase, however, radio emission
is lower than the current observational limit because the wind
velocity is as fast as $1000$ km s$^{-1}$ and the density of the
circumstellar medium is too tenuous to be observed even if the wind
mass loss rate is as large as $\sim 1 \times 10^{-6} M_\odot$
yr$^{-1}$ or more.  If the wind has stopped before the explosion, the
circumstellar matter is dominated by the low velocity wind from the
red-giant companion; then observations of radio emission would be
easier.  Detection of high velocity hydrogen feature from the strong
wind is also predicted.

\end{enumerate}

\acknowledgments
     We thank the referee, Mario Livio, for his helpful comments 
to improve the manuscript.  We are also grateful to Lev Yungelson 
for his kind comments and corrections on the manuscript.
This research has been supported in part by the Grant-in-Aid for
Scientific Research (05242102, 06233101, 08640321, 09640325) 
and COE research (07CE2002) of the Japanese Ministry of Education, 
Science, Culture, and Sports.

\appendix
\section{Specific angular momentum of winds}
     To obtain the relation between 
$v / a \Omega_{\rm orb}$ and 
$\ell_{\rm w}$ in equations
(\ref{specific_angular_momentum_wind}) and 
(\ref{separation_decrease_by_wind}), 
we  calculate many orbits of test particles 
moving under the Roche potential and Coriolis force, i.e.,
\begin{eqnarray}
{{d^2 x} \over {d t^2}} &=& 2 {{d y} \over {d t}} + x  
- {{1} \over {1+q}}{{x-x_1} \over {r_1^3}}
- {{q} \over {1+q}}{{x-x_2} \over {r_2^3}}, \cr
{{d^2 y} \over {d t^2}} &=& - 2 {{d x} \over {d t}} + y  
- {{1} \over {1+q}}{{y} \over {r_1^3}}
- {{q} \over {1+q}}{{y} \over {r_2^3}}, \cr
{{d^2 z} \over {d t^2}} &=&
- {{1} \over {1+q}}{{z} \over {r_1^3}}
- {{q} \over {1+q}}{{z} \over {r_2^3}},
\label{xyz-motion}
\end{eqnarray}
where $(x, y, z)$ is the position of the particle, 
$q= M_2/M_1$ the mass ratio, $r_1$ and $r_2$ are 
the distances from the primary and from the secondary 
to the test particle, respectively, and calculated from 
\begin{eqnarray}
r_1^2 &=& \left( x-x_1 \right)^2 + y^2 + z^2, \cr
r_2^2 &=& \left( x-x_2 \right)^2 + y^2 + z^2.
\label{distance-12}
\end{eqnarray}
Setting the center of gravity of the binary 
at the origin of the coordinates, we assume that 
the primary is located on the negative
side of the x-axis $(x_1, 0, 0)$ 
and the secondary is located on the positive side of
the x-axis $(x_2, 0, 0)$, i.e.,
\begin{eqnarray}
x_1 &=& - {{q} \over {1+q}}, \cr
x_2 &=& {{1} \over {1+q}}.
\label{position-12}
\end{eqnarray}
We also assume $a=1$, $G=1$, and $\Omega_{\rm orb}=1$ (or $M_1+M_2=1$) 
in our dimensionless form.
\par
     The primary (AGB star) blows spherically symmetric winds
with the initial velocity of $v_0$.  This process can be simulated 
by ejecting test particles from the primary surface 
at the radius which is much 
smaller than the inner critical Roche lobe, e.g., one tenth 
of the inner critical Roche lobe of the primary; i.e., we set
\begin{equation}
v = v_0, \mbox{\qquad at ~} r_1= 0.1 R_1^*.
\label{wind_initial_velocity}
\end{equation}
The trajectory of the wind may be approximated by
a trajectory of the particle with the same initial velocity and
position when the wind is supersonic and does not form a shock.  
Here, we assume the equatorial symmetry of the wind.
Dividing the primary surface into $64 \times 256$ parts, i.e., 
the azimuthal angle (from $\phi=0$ to 
$\phi=2 \pi$) into 256, ($\Delta \phi= 2 \pi/256$),
and the inclination angle (from $\theta=0$ to $\theta=\pi/2$) 
into 64 ($\Delta \theta= \pi/2/64$), we eject test particles 
from each center of the surface elements with the initial 
radial velocity of $v_0$.  We attach the mass loss rate of 
$v_0 \sin\theta_i \Delta \theta \Delta \phi /4 \pi$ to each
particle.
\par
      The radial component of the wind velocity $v_r$
near the inner critical Roche lobe is calculated from
\begin{equation}
v_r = 2 \sum_i {{1} \over {r_1}}
\left( {{d x_i} \over {d t}} (x_i-x_1) + 
{{d y_i} \over {d t}} y_i +{{d z_i} \over {d t}} z_i \right) 
{{\sin\theta_i \Delta \theta \Delta \phi} \over {4 \pi}},
\mbox{\qquad at ~} r_{1,i}= R_1^*,
\label{wind_velocity_Roche}
\end{equation}
where the position of each test particle is denoted 
by $(x_i, y_i, z_i)$ and the {\it radial} means the direction 
from the center of the primary to the test particle.
We estimate the average specific angular momentum 
of the test particles by
\begin{equation}
\ell_{\rm w} = 2 \sum_i \left( x_i^2+y_i^2 + x_i {{d y_i} \over {d t}}
- y_i {{d x_i} \over {d t}} y \right)
{{\sin\theta_i \Delta \theta \Delta \phi} \over {4 \pi}} 
 / 2 \sum_i {{\sin\theta_i \Delta \theta \Delta \phi} \over {4 \pi}},
\mbox{\qquad at ~} r_i= 10.
\label{specific_angular_momentum_test}
\end{equation}
It should be noted that some test particles are trapped 
to the secondary and never reach the radius of $r=10$ 
when the radial velocity is smaller than 
the orbital velocity, i.e., $v_r < a \Omega_{\rm orb} = 1$.
We do not include these particles in the calculation 
of the specific angular momentum of the wind in equation
(\ref{specific_angular_momentum_test}).
\par
      The integration of equation (\ref{xyz-motion}) is 
based on the second order leap-frog method.
Five cases of the mass ratio, $q=M_2/M_1= 3$, 2, 1, 
$1/2$, and $1/3$, are calculated for various initial 
velocity $v_0$.  The relation between the specific 
angular momentum $\ell_{\rm w}$ and the radial velocity
near the inner critical Roche lobe of the primary $v (\equiv v_r)$
is plotted in Figure \ref{angratio5}.
When the radial velocity of the wind is faster than twice the
orbital velocity, i.e., $v \gtrsim 2$ 
(or $v \gtrsim 2 a \Omega_{\rm orb}$), the specific angular
momentum is approximated by the limiting value of equation
(\ref{fast_wind_angular_momentum}) for very fast winds.
For wind velocities lower than 2, the specific angular 
momentum rapidly increases.  We find that the values are located
approximately on a quadratic line of
\begin{equation}
\ell_{\rm w} \approx 1.7 - 0.55 v^2,
\label{quadratic_wind_angular_momentum_appendix}
\end{equation} 
where the limiting case of
$\ell_{\rm w}=1.7$ for $v=0$ was obtained by
Nariai 
(1975) and
Nariai \& Sugimoto 
(1976) for a test particle simulation 
and Sawada et al. 
(1984) for a two-dimensional (equatorial plane) hydrodynamic 
simulation.
Thus, the specific angular momentum is approximated by 
\begin{equation}
\ell_{\rm w} \approx \max\left[ 1.7 - 0.55
\left( {{v} \over {a \Omega_{\rm orb}}} \right)^2,
\left( {{q} \over {1+q}} \right)^2 \right],
\label{wind_angular_momentum_appendix}
\end{equation} 
at least for these five different mass ratios, although 
three-dimensional hydrodynamic calculations 
should be done in order to obtain
a definite conclusion of the specific angular momentum of the winds.

\placefigure{angratio5}

%
%

\begin{deluxetable}{cccccc}
\footnotesize
\tablecaption{Numerical factors of mass-stripping effect
\label{tbl-1}}
\tablewidth{0pt}
\tablehead{
\colhead{$q$} & \colhead{$0.5$}   & \colhead{$1$}   
& \colhead{$2$} & \colhead{$3$}  & \colhead{$5$}
} 
\startdata
$g(q)$ & 0.013 & 0.019 & 0.025 & 0.030 & 0.036 \nl
$\ell_{\rm s}(q)$ & 0.144 & 0.025 & 0.006 & 0.044 & 0.105  \nl
\enddata
\end{deluxetable}

\begin{deluxetable}{lcccccc}
\footnotesize
\tablecaption{Three typical cases of SN Ia progenitor evolution
\label{tbl_initial_parameters}}
\tablewidth{0pt}
\tablehead{
\colhead{case} & \colhead{$M_{\rm WD,0}$} & \colhead{$M_{\rm RG,0}$}   
& \colhead{$P_0$} & \colhead{$\dot M_{\rm t}$} 
& \colhead{$\dot M_{\rm w}$}  & \colhead{$M_{\rm s}$} \cr
\colhead{} & \colhead{$(M_\odot)$} & \colhead{$(M_\odot)$}   
& \colhead{(day)} & \colhead{$(M_\odot$ yr$^{-1})$} 
& \colhead{$(M_\odot$ yr$^{-1})$}  & \colhead{$(M_\odot$ yr$^{-1})$}
} 
\startdata
P1 & 1.0 & 2.0 & 300 & $-8.7 \times 10^{-7}$ & $-4.8 \times 10^{-7}$ 
& $-2.0 \times 10^{-6}$ \nl
P2 & 1.0 & 1.6 & 300 & $-6.2 \times 10^{-7}$ & $-2.3 \times 10^{-7}$ 
& $-1.0 \times 10^{-6}$  \nl
P3 & 1.0 & 1.3 & 300 & $-5.2 \times 10^{-7}$ & $-1.3 \times 10^{-7}$ 
& $-5.4 \times 10^{-7}$  \nl
\enddata
\end{deluxetable}

\begin{deluxetable}{lcccccc}
\footnotesize
\tablecaption{Three states of immediate SN Ia progenitors
\label{tbl_three_states}}
\tablewidth{0pt}
\tablehead{
\multicolumn{1}{l}{case} &
\multicolumn{5}{c}{history} &
\multicolumn{1}{c}{SN Ia explosion} \nl
} 
\startdata
\multicolumn{1}{l}{P1} &
\multicolumn{1}{c}{WIND} &
\multicolumn{1}{c}{$\longrightarrow$} &
\multicolumn{1}{c}{WIND} &
\multicolumn{1}{c}{$\longrightarrow$} &
\multicolumn{1}{c}{WIND} &
\multicolumn{1}{c}{WIND} \nl
\multicolumn{1}{l}{P2} &
\multicolumn{1}{c}{WIND} &
\multicolumn{1}{c}{$\longrightarrow$} &
\multicolumn{1}{c}{WIND} &
\multicolumn{1}{c}{$\longrightarrow$} &
\multicolumn{1}{c}{SSS} &
\multicolumn{1}{c}{SSS} \nl
\multicolumn{1}{l}{P3} &
\multicolumn{1}{c}{WIND} &
\multicolumn{1}{c}{$\longrightarrow$} &
\multicolumn{1}{c}{SSS} &
\multicolumn{1}{c}{$\longrightarrow$} &
\multicolumn{1}{c}{RN} &
\multicolumn{1}{c}{RN} \nl
\enddata
\end{deluxetable}

\begin{deluxetable}{ccccccc}
\footnotesize
\tablecaption{Contraction factor $(\alpha_{\rm CE}=1)$
\label{tbl_contraction_factor}}
\tablewidth{0pt}
\tablehead{
\colhead{$M_{\rm WD,0}$} & \colhead{$M_{1,0}$}   & \colhead{$a_{\rm i}$} 
& \colhead{$a_{\rm f}$} & \colhead{$P_0$} & \colhead{$M_2$}  
& \colhead{contraction} \nl
\colhead{$(M_\odot)$} & \colhead{$(M_\odot)$}   & \colhead{$(R_\odot)$} 
& \colhead{$(R_\odot)$} & \colhead{(day)} & \colhead{$(M_\odot)$}  
& \colhead{factor}
} 
\startdata
$0.7$ & $3.20$ & $730-21,900$ & $64-1,920$ & $46-7,480$ & $1.0$ & $1/11$ \nl
$0.8$ & $4.48$ & $910-29,300$ & $44-1,420$ & $25-4,640$ & $1.0$ & $1/21$ \nl
$0.9$ & $5.60$ & $1,060-35,800$ & $36-1,220$ & $18-3,600$ & $1.0$ & $1/29$ \nl
$1.0$ & $6.63$ & $1,200-41,700$ & $32-1,120$ & $15-3,070$ & $1.0$ & $1/37$ \nl
$1.1$ & $7.58$ & $1,340-47,200$ & $30-1,060$ & $13-2.740$ & $1.0$ & $1/45$ \nl
$1.2$ & $8.48$ & $1,460-52,400$ & $28-1,020$ & $12-2.530$ & $1.0$ & $1/52$ 
\enddata
\end{deluxetable}

\begin{deluxetable}{cccccc}
\footnotesize
\tablecaption{Realization frequency of SNe Ia ($\eta_{\rm eff}=1$)
\label{tbl_realizaton_frequency}}
\tablewidth{0pt}
\tablehead{
\colhead{$M_{\rm WD,0}$} & \colhead{$\Delta \log A$}   
& \colhead{$M_A$} & \colhead{$M_B$} & \colhead{$\Delta q$} 
& \colhead{$\nu_{\rm WD}$} \nl  
\colhead{$(M_\odot)$} & \colhead{}   & \colhead{$(M_\odot)$} 
& \colhead{$(M_\odot)$} & \colhead{} & \colhead{(yr$^{-1}$)}  
} 
\startdata
$0.8-0.9$ & $0.6 \cdot 2/3$ & $4.48$ & $5.60$ & $0.37$ 
& $0.0006$ \nl
$0.9-1.0$ & $1.0 \cdot 2/3$ & $5.60$ & $6.63$ & $0.36$ 
& $0.0006$ \nl
$1.0-1.1$ & $1.5 \cdot 2/3$ & $6.63$ & $7.58$ & $0.36$ 
& $0.0005$ 
\enddata
\end{deluxetable}

\begin{deluxetable}{cccccc}
\footnotesize
\tablecaption{Realization frequency of SNe Ia ($\eta_{\rm eff}=0.3$)
\label{tbl_realization_0.33}}
\tablewidth{0pt}
\tablehead{
\colhead{$M_{\rm WD,0}$} & \colhead{$\Delta \log A$}   
& \colhead{$M_A$} & \colhead{$M_B$} & \colhead{$\Delta q$} 
& \colhead{$\nu_{\rm WD}$} \nl  
\colhead{$(M_\odot)$} & \colhead{}   & \colhead{$(M_\odot)$} 
& \colhead{$(M_\odot)$} & \colhead{} & \colhead{(yr$^{-1}$)}  
} 
\startdata
$0.9-1.0$ & $0.7 \cdot 2/3$ & $5.60$ & $6.63$ & $0.21$ 
& $0.0002$ \nl
$1.0-1.1$ & $1.1 \cdot 2/3$ & $6.63$ & $7.58$ & $0.23$ 
& $0.0003$ \nl
$1.1-1.2$ & $1.8 \cdot 2/3$ & $7.58$ & $8.48$ & $0.23$ 
& $0.0003$ 
\enddata
\end{deluxetable}

%
%
%
%

\newpage

\begin{figure}
\plotone{channel.eps}
\caption{
An illustration of the symbiotic channel to Type Ia supernovae.
\label{channel}}
\end{figure}


\begin{figure}
\plotone{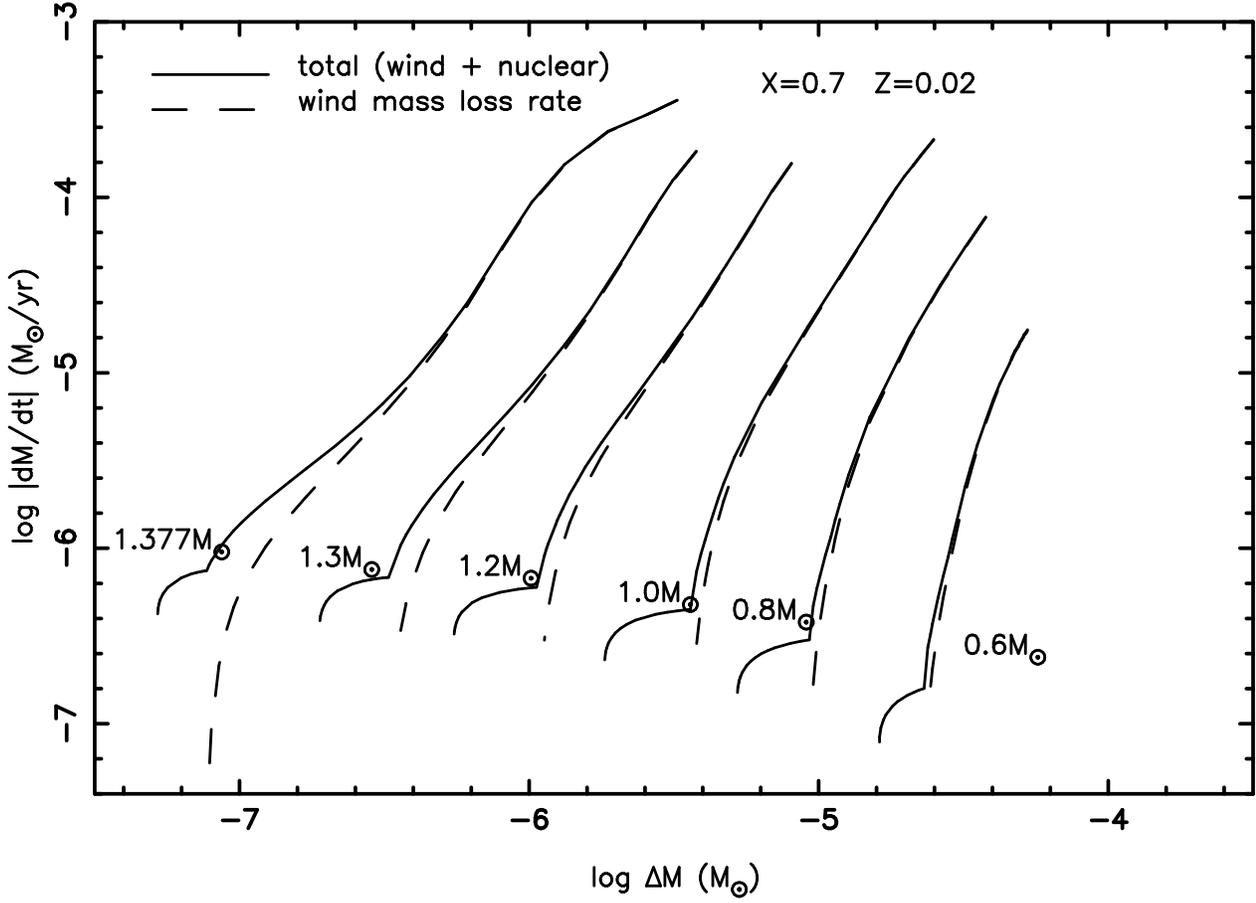}
\caption{
Wind mass loss rate $\dot M_{\rm wind}$ (dashed line) 
and the total mass decreasing rate of the hydrogen-rich envelope 
on the white dwarf $\dot M_{\rm nuc} + \dot M_{\rm wind}$
(solid line), i.e., nuclear burning rate $\dot M_{\rm nuc}$ 
plus wind mass loss rate $\dot M_{\rm wind}$, 
are plotted against the envelope mass $\Delta M$ for WDs with 
masses of $0.6 M_\odot$, $0.8 M_\odot$, $1.0 M_\odot$, $1.2 M_\odot$, 
$1.3 M_\odot$, and $ 1.377 M_\odot$.
White dwarf mass is attached to each line.
The metallicity is the solar value of $Z= Z_\odot = 0.02$.  
There exists only a static (no wind) solution below the break 
of each solid line while there exists only a wind solution
above the break for a given envelope mass of $\Delta M$.
The optically thick winds blow when the mass transfer rate
from the companion star 
$|\dot M_2| = |\dot M_{\rm nuc} + \dot M_{\rm wind}|$ 
exceeds $\dot M_{\rm cr} = 0.75 \times 10^{-6} (M_{\rm WD}/M_\odot - 0.4)
M_\odot$ yr$^{-1}$.  
\label{dmdtenvx70z02}}
\end{figure}

\begin{figure}
\plotone{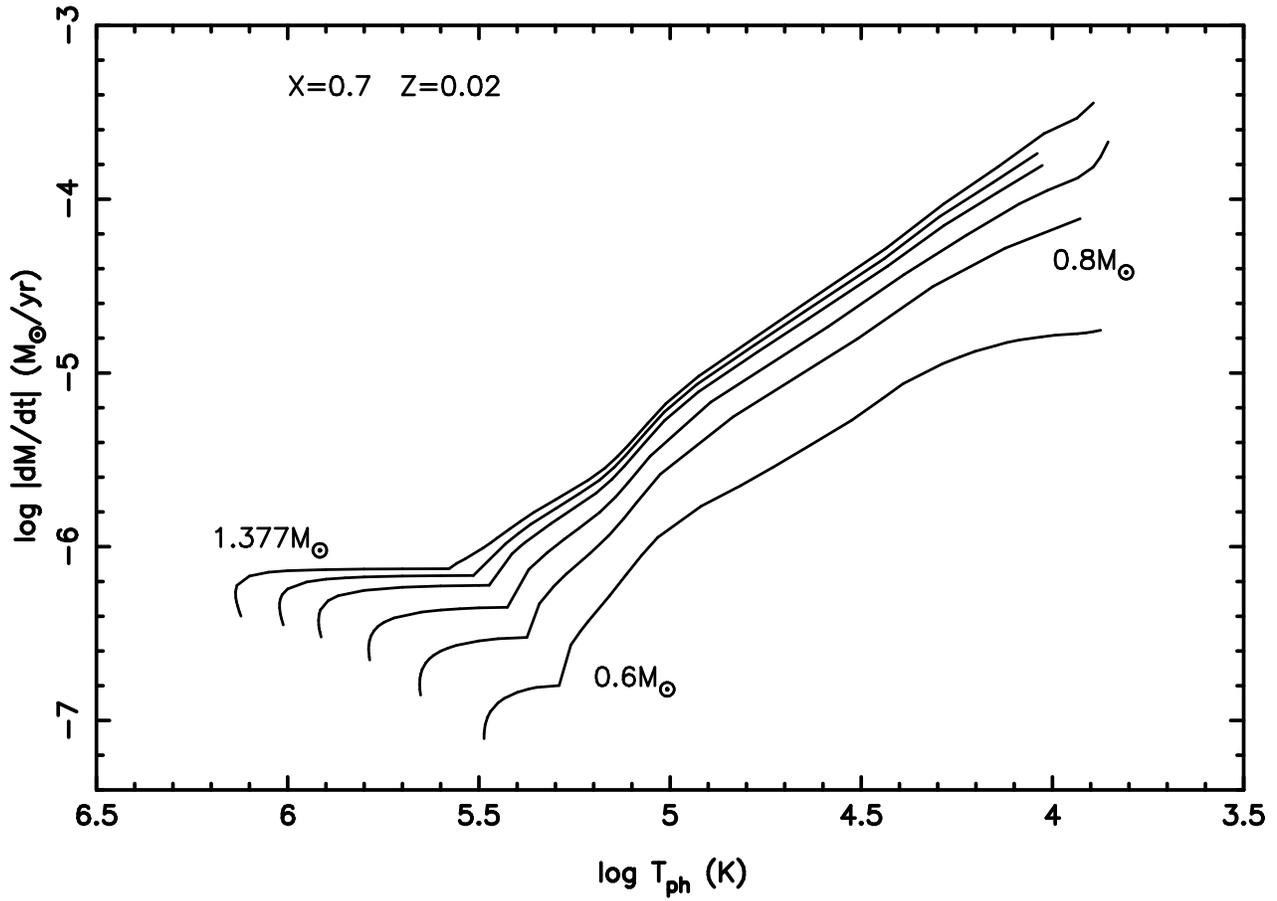}
\caption{
Photospheric temperature $T_{\rm ph}$ is 
plotted against the decreasing rate of 
the envelope mass $\dot M_{\rm nuc} + \dot M_{\rm wind}$
for WDs with masses of 
$0.6 M_\odot$, $0.8 M_\odot$, $1.0 M_\odot$, 
$1.2 M_\odot$, $1.3 M_\odot$, and 
$1.377 M_\odot$.   Same envelope models as in Fig. 2. 
Optically thick winds blow when 
$\log T_{\rm ph} ~({\rm K}) \lesssim 5.3$.
\label{dmdttphx70z02}}
\end{figure}

\begin{figure}
\plotone{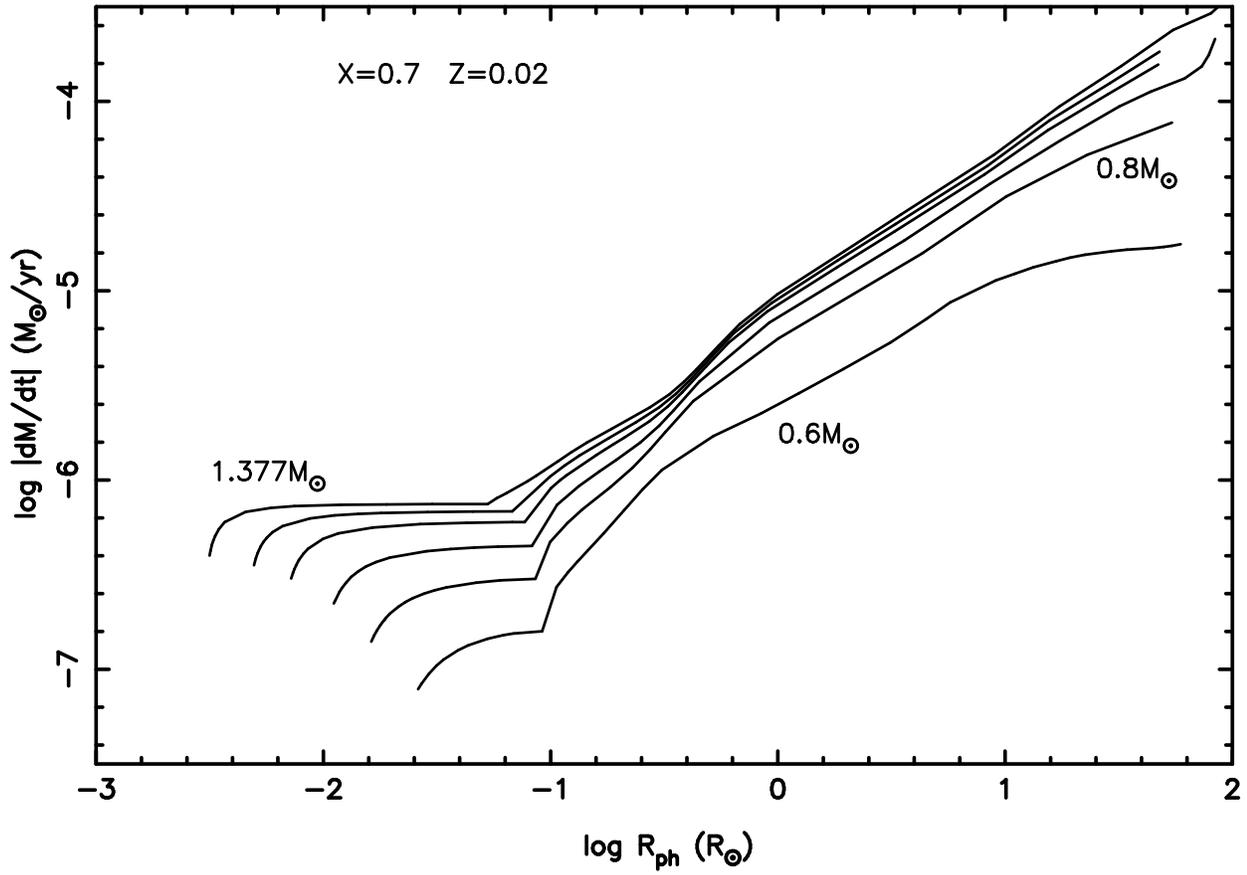}
\caption{
Photospheric radius $R_{\rm ph}$ is plotted 
against the decreasing rate of 
the envelope mass $\dot M_{\rm nuc} + \dot M_{\rm wind}$
for WDs with masses of 
$0.6 M_\odot$, $0.8 M_\odot$, $1.0 M_\odot$, 
$1.2 M_\odot$, $1.3 M_\odot$, and 
$1.377 M_\odot$.   Same envelope models as in Fig. 2. 
Optically thick winds blow when $R_{\rm ph} \gtrsim 0.1 R_\odot$.
\label{dmdtrphx70z02}}
\end{figure}

\begin{figure}
\plotone{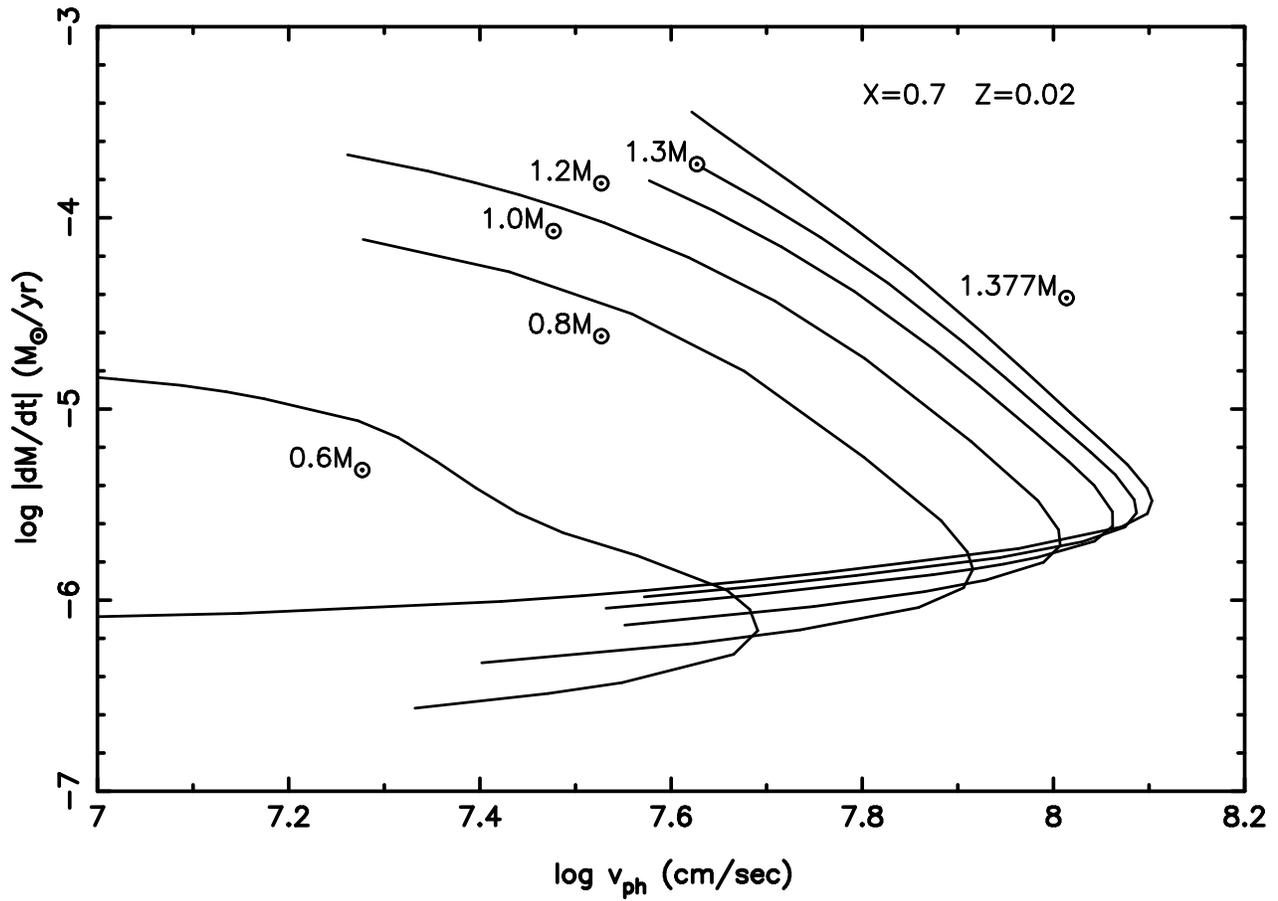}
\caption{
Photospheric velocity $v_{\rm ph}$ is
plotted against the decreasing rate of 
the envelope mass $\dot M_{\rm nuc} + \dot M_{\rm wind}$
for WDs with masses of 
$0.6 M_\odot$, $0.8 M_\odot$, $1.0 M_\odot$, 
$1.2 M_\odot$, $1.3 M_\odot$, and 
$1.377 M_\odot$.   Same models as in Fig. 2. 
\label{dmdtvelx70z02}}
\end{figure}

\begin{figure}
\plotone{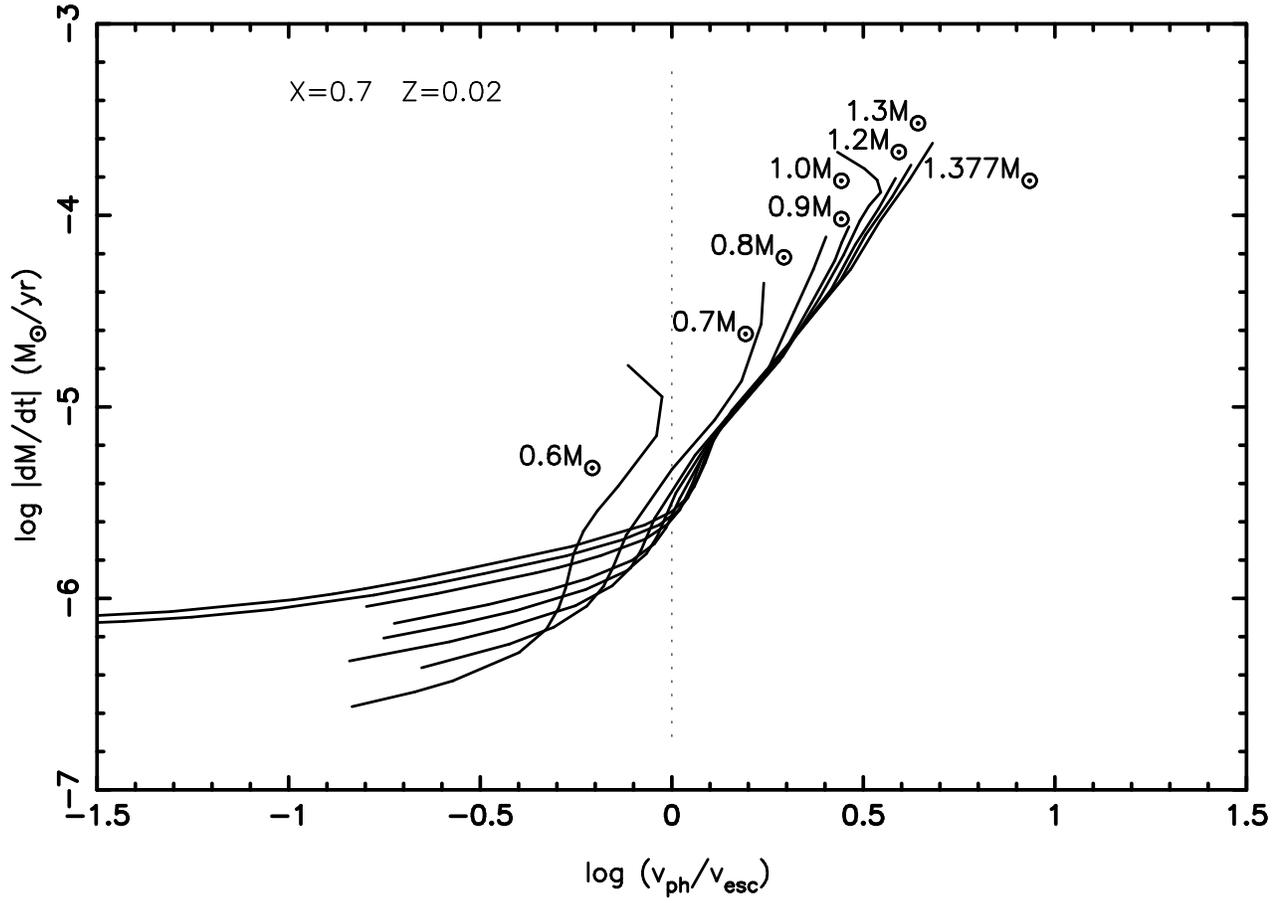}
\caption{
Ratio of the photospheric velocity to the escape 
velocity there $v_{\rm ph} / v_{\rm esc}$ 
is plotted against the decreasing rate of 
the envelope mass $\dot M_{\rm nuc} + \dot M_{\rm wind}$
for WDs with masses of 
$0.6 M_\odot$, $0.7 M_\odot$, $0.8 M_\odot$, $0.9 M_\odot$, 
$1.0 M_\odot$, $1.2 M_\odot$, $1.3 M_\odot$, and 
$1.377 M_\odot$.   Same models as in Fig. 2, 
but $0.7 M_\odot$ and $0.9 M_\odot$ are added. 
We regard the wind as ``strong'' when the photospheric 
velocity exceeds the escape velocity there, i.e.,
$v_{\rm ph} > v_{\rm esc}$.  If not, it is regarded as
``weak.'' 
\label{dmdtescx70z02}}
\end{figure}

\begin{figure}
\plotone{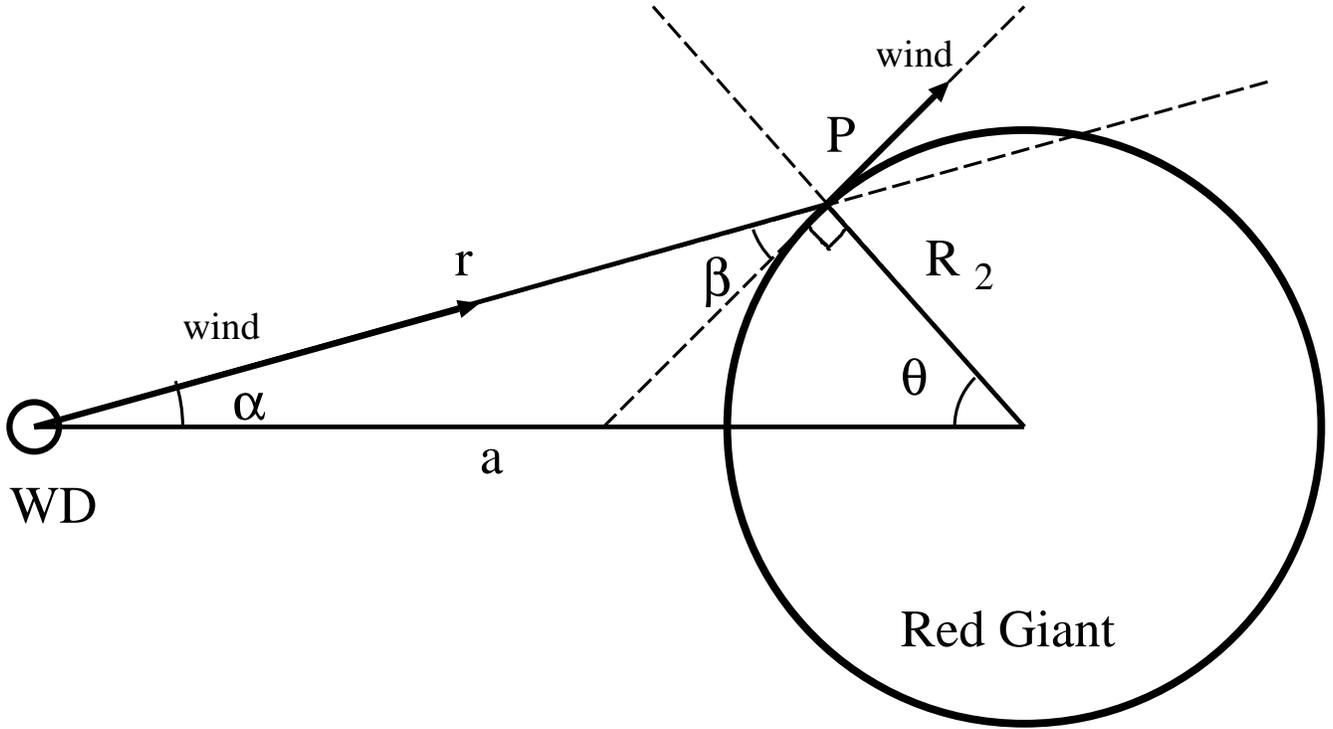}
\caption{
Fast winds from the white dwarf (WD) 
collide with the surface of the red-giant and 
strip mass from the red-giant.  The normal component of the wind 
velocity to the red-giant surface is dissipated by forming a shock 
and it heats up the surface.  
Then, a part of the surface mass is ablated and blown in the wind. 
\label{collision}}
\end{figure}

\begin{figure}
\plotone{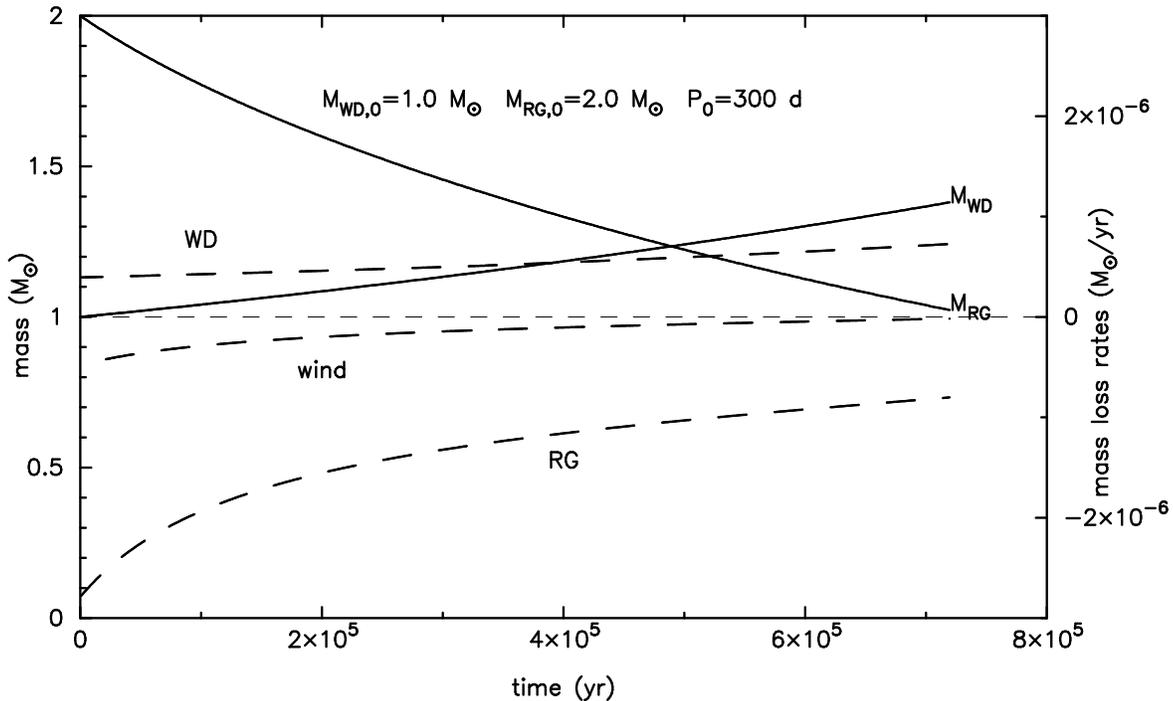}
\caption{
Time evolution of an SN Ia progenitor system for 
case P1 (explosion during the wind phase).  
The initial parameters are shown at the top of the figure.
The white dwarf mass increases 
to $1.38 M_\odot$ and explodes as an SN Ia at
$t= 7.2 \times 10^5$ yr.  The solid
lines show the masses of the white dwarf ($M_{\rm WD}$) 
and the red-giant companion ($M_{\rm RG}$).
The dashed lines show, from top to bottom, 
the net mass accretion rate onto the white
dwarf, the wind mass loss rate, and the mass decreasing rate 
of the red-giant companion, respectively.
\label{evolution_wind}}
\end{figure}

\begin{figure}
\plotone{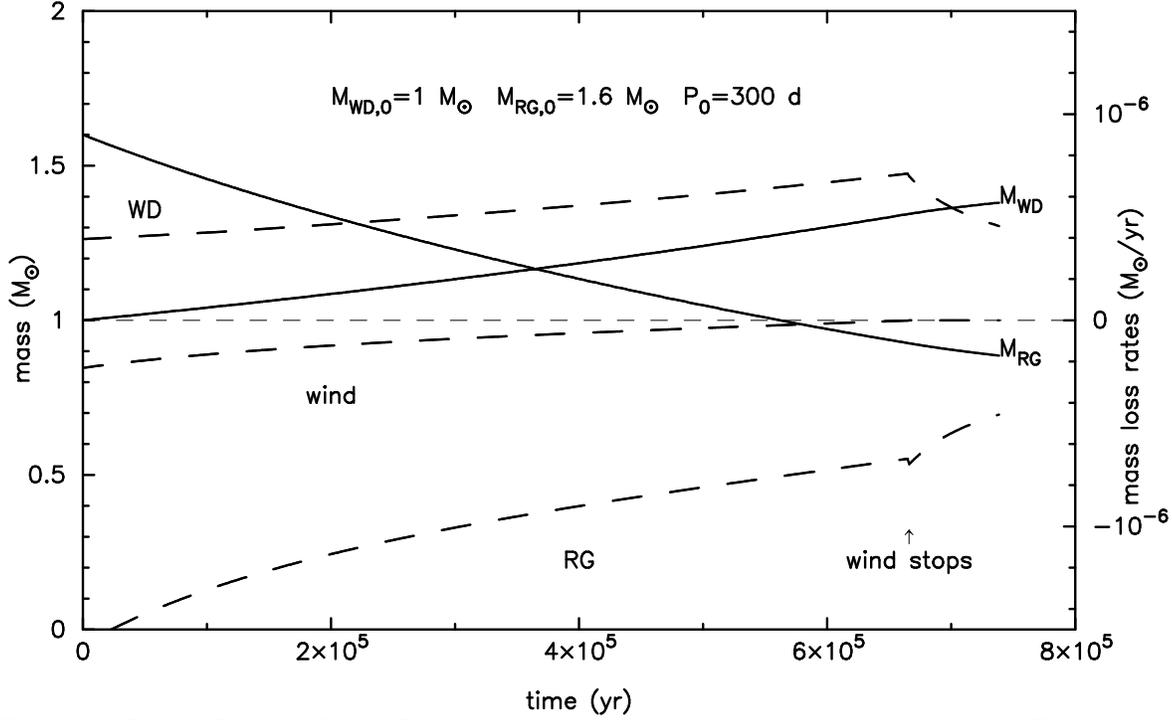}
\caption{
Same as Fig. 8 but for case P2 (explosion during the steady hydrogen 
shell burning phase).
The strong wind stops at the time indicated by an arrow.
The white dwarf mass increases 
to $1.38 M_\odot$ and explodes as an SN Ia at
$t= 7.4 \times 10^5$ yr.
\label{evolution_nuc}}
\end{figure}

\begin{figure}
\plotone{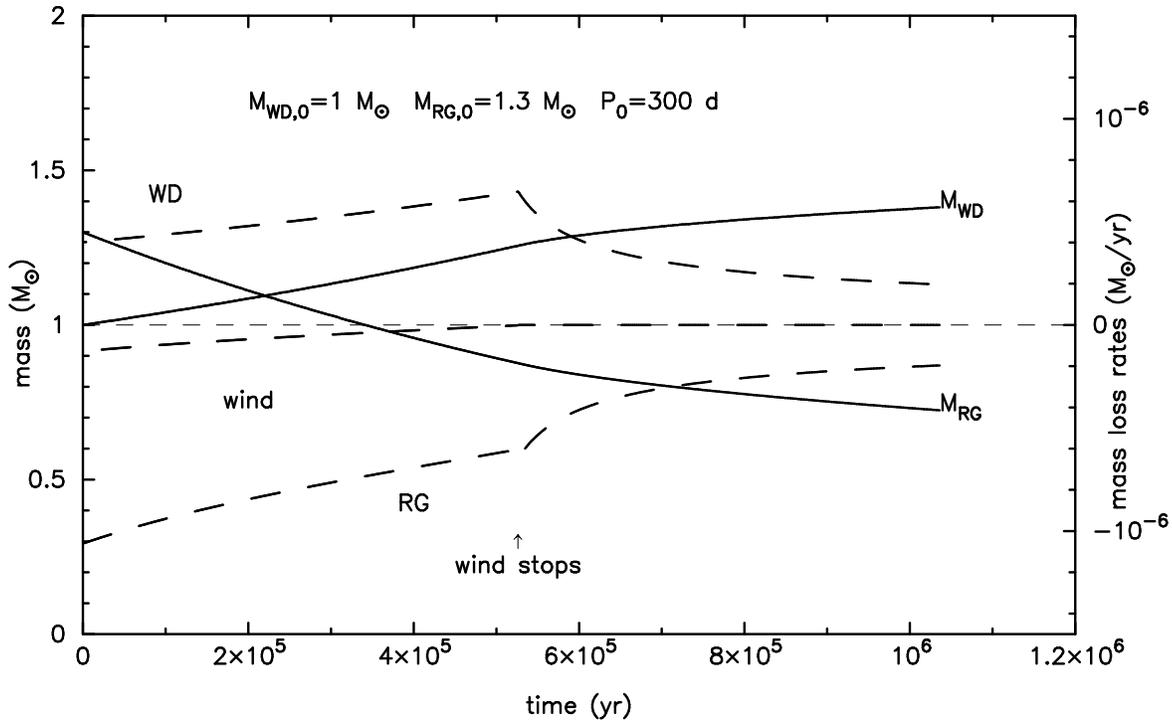}
\caption{
Same as Fig. 8 but for case P3 (explosion during the very weak shell 
flash phase).
The strong wind stops at the time indicated by an arrow.
The hydrogen shell burning becomes unstable to trigger very weak
shell flashes at $t=6.5 \times 10^5$ yr.
The white dwarf mass increases 
to $1.38 M_\odot$ and explodes as an SN Ia at
$t= 1.04 \times 10^6$ yr.
\label{evolution_unstable}}
\end{figure}

\begin{figure}
\plotone{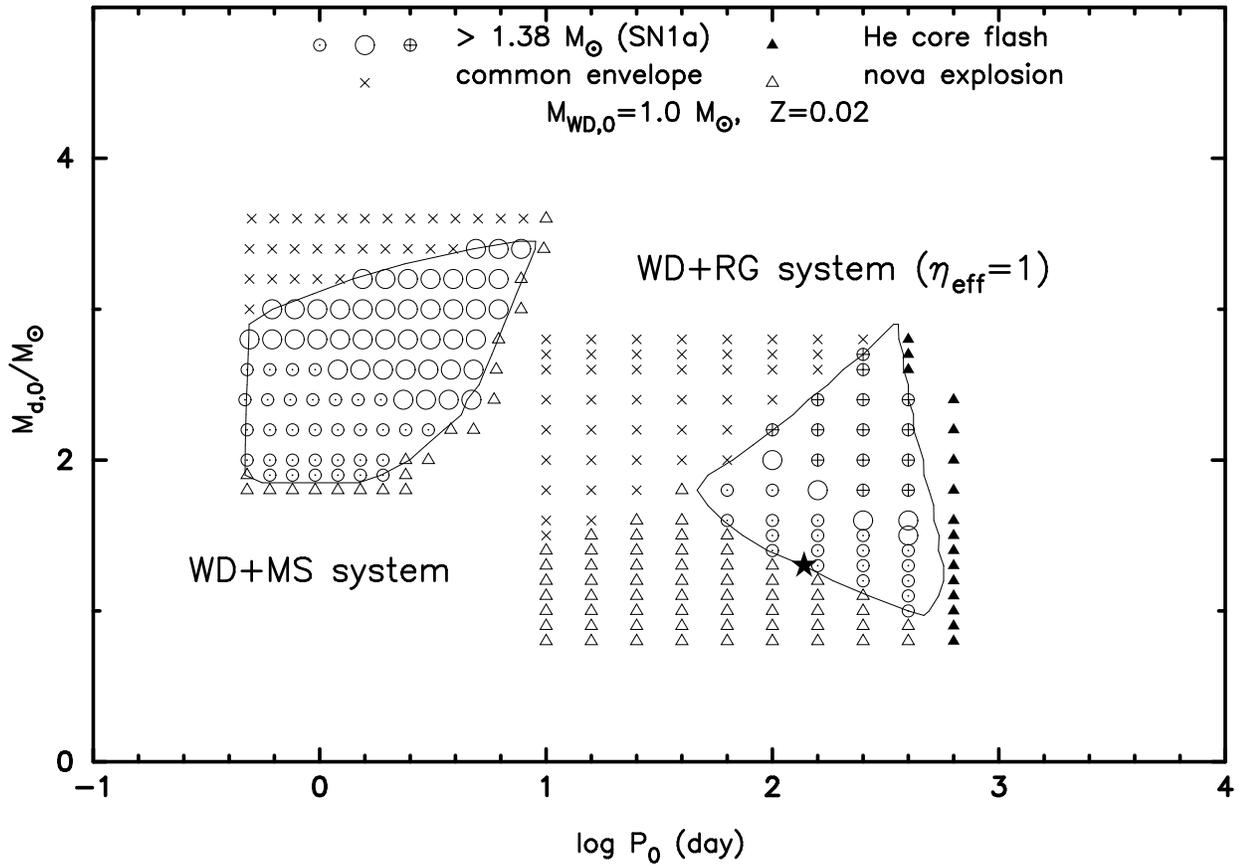}
\caption{
Final outcome of close binary evolution 
in the $\log P_0 - M_{\rm d,0}$ plane.  Here, $M_{\rm d,0}$ is 
the donor mass, $M_{\rm MS,0}$ or $M_{\rm RG,0}$. 
Final outcome is either 
an unstable mass transfer ($H_4(q) > 0$) at the beginning
(forming a common envelope; denoted by $\times$), 
or an SN Ia explosion (denoted by $\oplus$, $\bigcirc$, or $\odot$) 
or a nova (denoted by a open triangle), 
or a central helium flash (denoted by a filled triangle).  
$\oplus$: wind phase at SN Ia explosion (P1). $\bigcirc$: wind stops 
before SN Ia explosion but the mass transfer rate is still 
high enough to keep steady hydrogen shell burning, 
i.e., $|\dot M_{\rm t}| > \dot M_{\rm st}$ (P2). 
$\odot$: wind stops before SN Ia explosion and the mass transfer 
rate is decreasing between $ \dot M_{\rm low} <
|\dot M_{\rm t}| < \dot M_{\rm st}$ at SN Ia explosion (P3). 
The region producing an SN Ia is bounded by a solid line. 
The left/right region corresponds to the WD+MS (compact)/WD+RG (wide) 
system, respectively.  
A star mark ($\star$) denotes an initial position of T CrB in the 
WD+RG systems.
\label{zams10}}
\end{figure}

\begin{figure}
\plotone{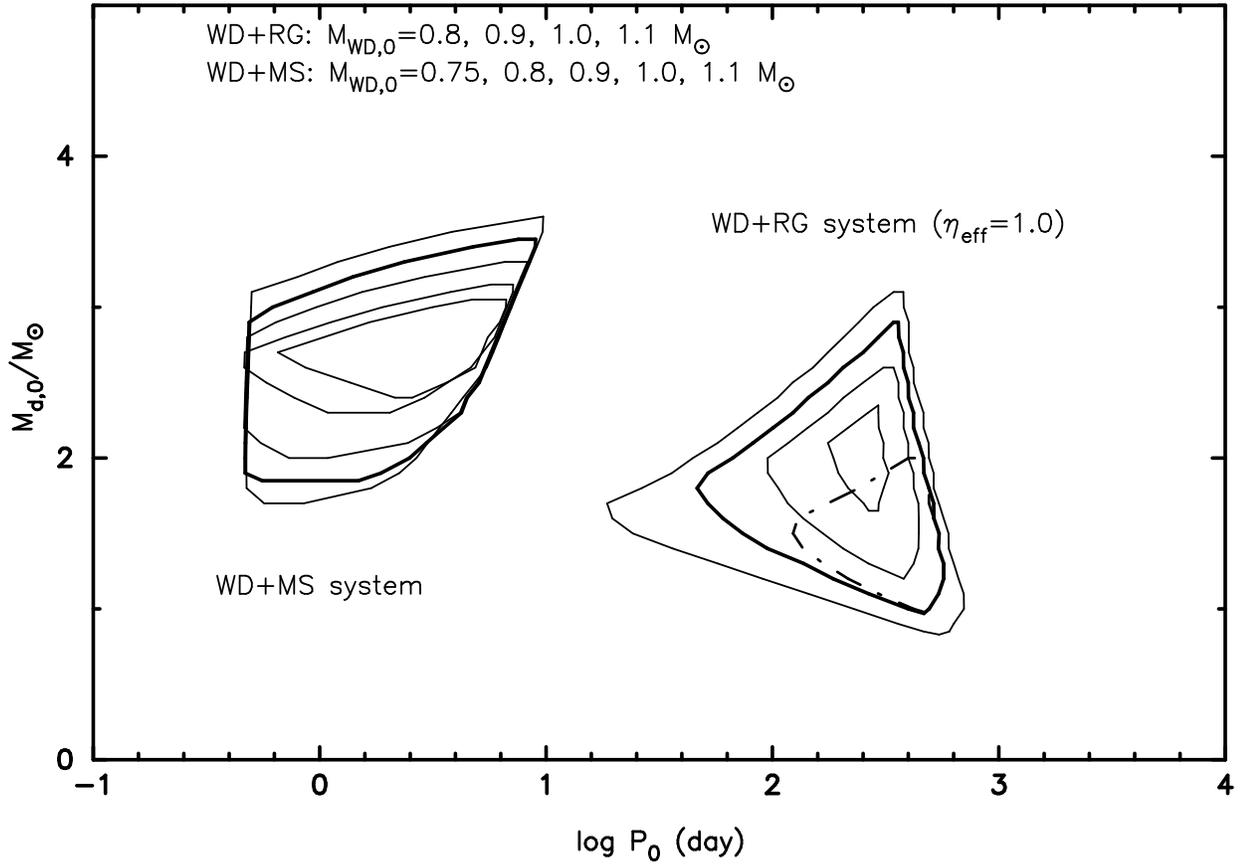}
\caption{
The region to produce SNe Ia 
in the $\log P_0 - M_{\rm d,0}$ plane for five initial white 
dwarf masses of $0.75 M_\odot$, $0.8 M_\odot$, 
$0.9 M_\odot$, $1.0 M_\odot$ (heavy solid line), 
and $1.1 M_\odot$.  The region of $M_{\rm WD,0}= 0.7 M_\odot$ 
almost vanishes for both the WD+MS and WD+RG systems, and 
the region of $M_{\rm WD,0}= 0.75 M_\odot$ vanishes for 
the WD+RG system.  Here, we assume the stripping efficiency of 
$\eta_{\rm eff}=1$.  For comparison, we show only 
the region of $M_{\rm WD,0}= 1.0 M_\odot$ for a much lower 
efficiency of  $\eta_{\rm eff}=0.3$ by a dash-dotted line.
\label{ztotreg100}}
\end{figure}

\begin{figure}
\plotone{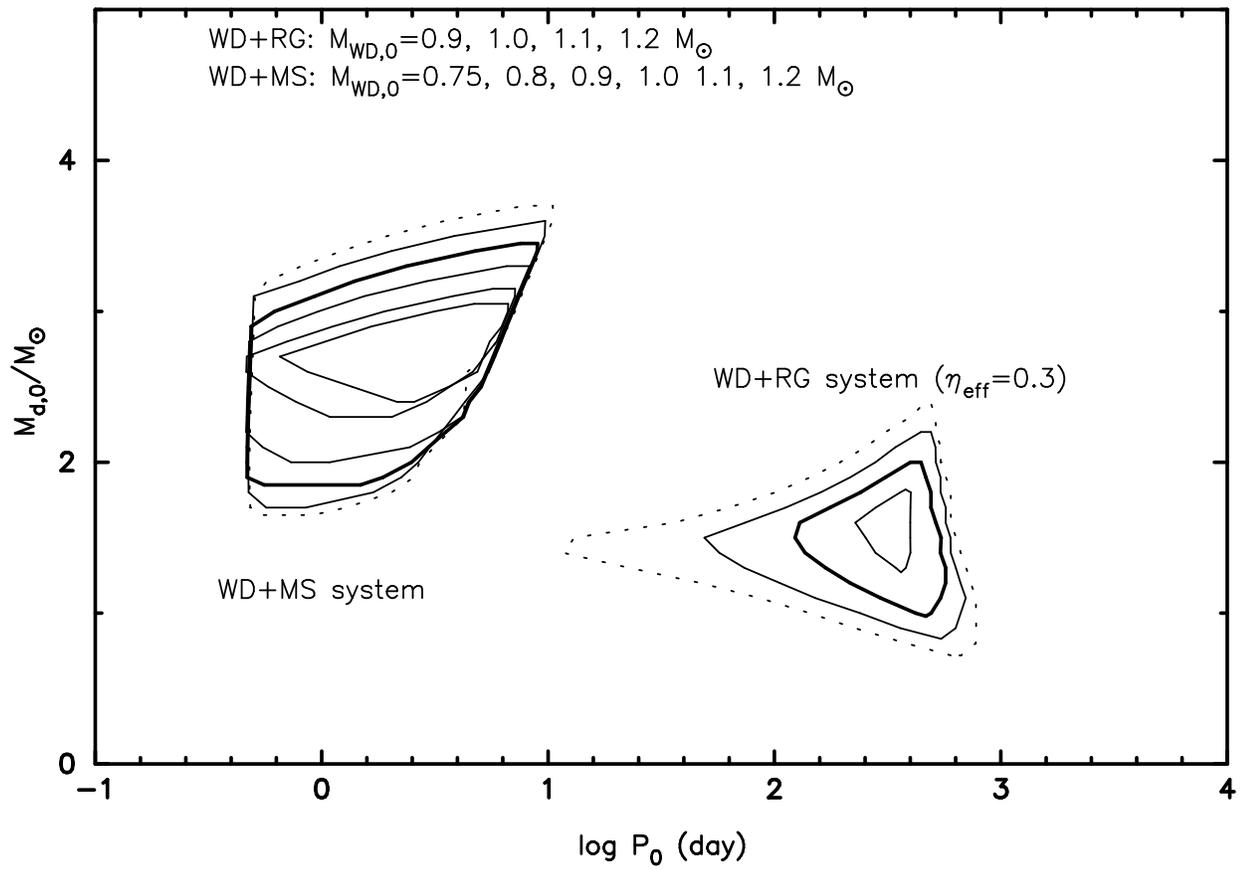}
\caption{
Same as Fig. 12 but for the much lower mass-stripping 
efficiency of $\eta_{\rm eff}=0.3$.  We add the region of 
$M_{\rm WD,0}= 1.2 M_\odot$ both for the WD+MS/WD+RG systems
(dotted lines).  The region of $M_{\rm WD,0}= 0.8 M_\odot$ vanishes for 
the WD+RG system.
\label{ztotreg030}}
\end{figure}

\begin{figure}
\plotone{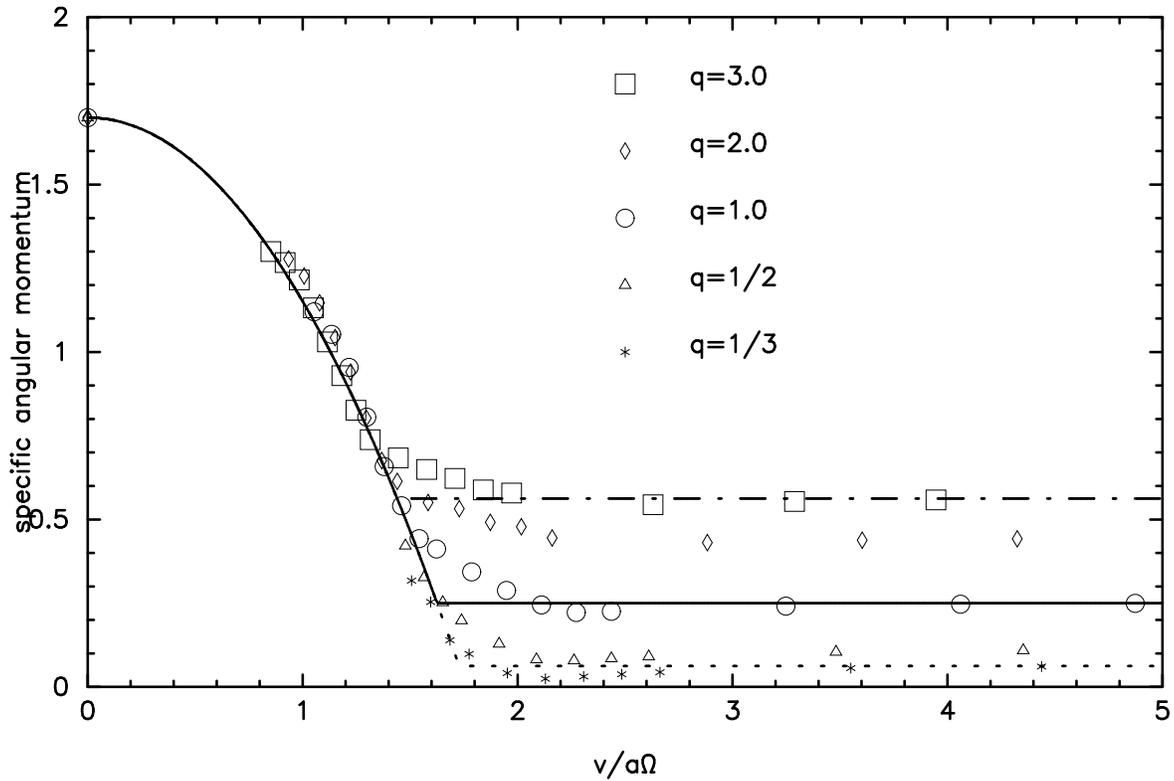}
\caption{
Specific angular momentum of wind is plotted against 
the outflowing velocity near the inner critical Roche lobe.
The specific angular momentum and the outflowing velocity are
measured in units of $a^2\Omega$ and $a\Omega$, respectively, 
where $a$ is the separation and $\Omega$ is the
orbital angular velocity.  
Five cases of the mass ratio are examined, i.e., 
$q=M_2/M_1= 3$, 2, 1, $1/2$, and $1/3$.  It is assumed that 
the wind blows from the primary.   
The limiting case of $\ell_{\rm w}=1.7$ for $v=0$ is taken 
from Nariai (1975), Nariai \& Sugimoto (1976), 
and Sawada et al. (1984).
\label{angratio5}}
\end{figure}

%

\end{document}